\documentclass[9pt,shortpaper,twoside,web]{IEEEtran}
\usepackage[ruled,noline]{algorithm2e}
\usepackage{algorithmic}
\usepackage{setspace}
\usepackage{amssymb}
\usepackage{amsmath}
\usepackage{latexsym}
\usepackage{framed}
\usepackage{color, soul}
\usepackage{xargs}
\usepackage{tikz}
\usepackage{todonotes}
\usepackage{multirow} 
\usepackage{verbatim}

\usepackage{xcolor} 

\usepackage{array}
\usepackage{mdwtab}
\usepackage{eqparbox}
\usepackage{url}

\usepackage{graphicx}

\begin{document}

\title{Unpaired cross-modality educed distillation (CMEDL) for medical image segmentation}

\author{Jue Jiang, Andreas Rimner, Joseph O. Deasy, and Harini Veeraraghavan
\thanks{This work was partially supported by the MSK Cancer Center support grant/core grant P30 CA008748 and NCI R01 CA198121.}
\thanks{J. Jiang, J.O. Deasy, and H. Veeraraghavan are all with the Department of Medical Physics, Memorial Sloan Kettering Cancer Center, NY (e-mail: jiangj1@mskcc.org; deasyJ@mskcc.org; veerarah@mskcc.org). }
\thanks{A. Rimner is with Department of Radiation Oncology, Memorial Sloan Kettering Cancer Center, NY(e-mail: rimnera@mskcc.org).}
}

\maketitle

\begin{abstract}
Accurate and robust segmentation of lung cancers from CT, even those located close to mediastinum, is needed to more accurately plan and deliver radiotherapy and to measure treatment response. Therefore, we developed a new cross-modality educed distillation (CMEDL) approach, using unpaired CT and MRI scans, whereby an \textcolor{black}{informative} teacher MRI network guides a student CT network to extract features that signal the difference between foreground and background. Our contribution eliminates two requirements of distillation methods: (i) paired image sets by using an image to image (I2I) translation and (ii) pre-training of the teacher network with a large training set by using concurrent training of all networks. Our framework uses an end-to-end trained unpaired I2I translation, teacher, and student segmentation networks. Architectural flexibility of our framework is demonstrated using 3 segmentation and 2 I2I networks. Networks were trained with 377 CT and 82 T2w MRI from different sets of patients, with independent validation (N=209 tumors) and testing (N=609 tumors) datasets. \textcolor{black}{Network design, methods to combine MRI with CT information, distillation learning under informative (MRI to CT), weak (CT to MRI) and equal teacher (MRI to MRI), and ablation tests were performed.} Accuracy was measured using Dice similarity (DSC), surface Dice (sDSC), and Hausdorff distance at the 95$^{th}$ percentile (HD95). The CMEDL approach was significantly (p $<$ 0.001) more accurate \textcolor{black}{(DSC of 0.77 vs. 0.73) than non-CMEDL methods with an informative teacher for CT lung tumor, with a weak teacher (DSC of 0.84 vs. 0.81) for MRI lung tumor, and with equal teacher (DSC of 0.90 vs. 0.88) for MRI multi-organ segmentation. CMEDL also reduced inter-rater lung tumor segmentation variabilities}.

\end{abstract}


	\section{Introduction}

	A key unmet need for accurate radiotherapy planning and treatment response assessment is robust and automated segmentation of lung cancers, including those abutting the mediastinum\cite{whitfield2013automated}. The low soft-tissue contrast on standard-of-care computed tomography (CT) presents a challenge to obtain robust and reproducible segmentations, needed for more precise image-guided treatments.
	\par
	Deep learning lung tumor segmentation methods\cite{jiang2019,pang2019fast,zhao2018tumor} already outperform non-deep learning methods\cite{tan2013segmentation}. Improved accuracies have been achieved through careful pre-processing to focus the algorithm towards slices containing tumor\cite{HossainICASSP19}, or searching only within lung parenchyma\cite{Hu2021TII}, and by using shape priors combined with three different views\cite{ByunSPIE2020}. However, such methods can be less reliable for tumors invading into the mediastinum or the chestwall, because pre-processing to use only the lung parenchyma can also exclude the tumors, and tumors invading into soft tissue often have different shapes than solid tumors encased within the lung tissue. Residual combination of features\cite{jiang2019,Zhang2020} alleviate the limitations of afore-mentioned methods, but their accuracies are less promising for mediastinal tumors. 
	\par
	Recent works\cite{leiPMB2020,FuMedPhys2020} used generative adversarial works that combined MRI and CT to derive a CT representation with better soft tissue contrast and improved organs segmentation accuracies. This was accomplished by synthesizing pseudo MRI (pMRI) from CT\cite{leiPMB2020} as well as by combining pMRI with CT\cite{FuMedPhys2020}. However, such methods require spatially accurate synthesis of pMRI, which is difficult to achieve in practical settings. Hence, we developed a distillation learning approach, where \textcolor{black}{MRI is used to guide the extraction of informative high level CT features during training. Once trained, MRI is not required for segmentation.}
	\par
	\textcolor{black}{Similar to the works in\cite{dou2020,Li_Yu_Wang_Heng_2020,KangMICCAI2020}, we performed unpaired distillation-based segmentation using different sets of CT and MR images. We also performed concurrent training of teacher and student networks, shown to be feasible for medical image segmentation\cite{dou2020,Li_Yu_Wang_Heng_2020,KangMICCAI2020}, and which obviates the need for pre-training the teacher network with large datasets\cite{hinton2015distilling,gupta2016cross,Kats2019}.}
	\par
	However, unlike\cite{Li_Yu_Wang_Heng_2020,KangMICCAI2020,hinton2015distilling}, which match the outputs of the teacher and student networks to perform distillation, we used "hint losses"\cite{romero2014fitnets} and match the intermediate features between the student and teacher networks. Concretely, high-level task features extracted by the teacher network from a synthesized modality (e.g. pseudo MRI corresponding to a CT image processed by the student network) is matched with the same features computed by the student network (e.g. from CT image). Using hint losses instead of matching output segmentations allows for segmentation variability in the two modalities, common due to differences in the tissue visualizations. This approach is thus different from\cite{Li_Yu_Wang_Heng_2020}, which trained the teacher (with pseudo target) and student (with target) networks with the same modalities and used both networks during testing to provide an ensemble segmentation. Our approach on the other hand only requires the student network during testing for generating the output segmentation, which requires smaller memory and fewer computations. We call our approach cross-modality educed distillation learning (CMEDL, pronounced "C-medal"). We also studied distillation under the setting of equally informative or ''equal'' teacher for same modality distillation (e.g. T1w MRI vs. T2w MRI) and weak teacher (CT to MRI) distillation-based segmentation.

    Our CMEDL framework consists of a cross-modality image-to-image translation (I2I) and concurrently trained teacher (MRI) and student (CT) segmentation networks. The I2I network allows for training with unpaired image sets by synthesizing corresponding pMRI images for knowledge distillation. 
	Our contributions are:
	\begin{itemize}
		\item An unpaired cross-modality distillation-based segmentation framework. Our default approach uses an \underline{informative} teacher (MRI with higher soft-tissue contrast) to guide the student CT network to extract features that signal the difference between foreground and background. 
		
		\item We also studied the performance of this framework with \underline{uninformative or weak teacher} for cross-modality (i.e. using CT for MRI segmentation) and \underline{equal teacher} for same modality but different contrast (T1w to T2w MRI and vice-versa) distillation.
		\item An architecture independent framework. We demonstrate feasibility using three different segmentation and two I2I networks. We discuss the tradeoffs and complexities in using different I2I networks for distillation.
		\item A concurrent mutual distillation framework that obviates the need for pre-training teacher network with large labeled datasets. We also studied distillation learning by training a teacher (MRI) network without MRI expert-segmentations.
		\item  We performed extensive analysis of accuracy under various conditions of using pMRI information and ablation experiments. 
	\end{itemize}
	
	Our paper substantially improves our previously published work on lung tumor segmentation\cite{jiang2019MICCAI}, with significant extensions and the following improvements: \textcolor{black}{(a) distillation learning using informative, uninformative or weak, and equally informative teacher modality, (b) same modality distillation with different contrasts (T1w to T2w MRI and vice versa) applied to a different problem of abdominal organs segmentation, (c) improved distillation framework with deeper multiple resolution residual network (MRRN) segmentation network\cite{jiang2019}, which significantly improves accuracy, (d) analysis of I2I synthesis accuracy and it's impact on segmentation accuracy with an additional and newer variational auto-encoder\textcolor{black}{\cite{lee2018diverse}} network, (e) analysis and discussion of tradeoffs with respect to accuracy and computational/training requirements in the choice of the segmentation and I2I networks used in CMEDL framework, (f) improved and more detailed explanation of the distillation framework, framework description with improved figures for network architectures, and improved notations. (g) We also provide: analysis on an enlarged test set of 609 patients (333 was used previously\cite{jiang2019MICCAI}); experiments to evaluate distillation learning without any expert-segmented data for the teacher network; compared to the recent method\cite{Li_Yu_Wang_Heng_2020}; accuracy evaluations using various strategies for incorporating pMRI information; study of why improved accuracy is achieved with distillation by using unsupervised clustering of foreground and background features with and without CMEDL method; inter-rater robustness evaluation; and ablation tests to study the design choices and loss functions to better inform the operating conditions and limitations of our approach.}
	
	\section{Related works}
	\subsection{Distillation learning as model compression}
	Distillation learning was initially developed as an approach for knowledge compression applied to object classification\cite{hinton2015distilling}, whereby simpler models with fewer parameters were extracted from a pre-trained high-capacity teacher network. Distillation was accomplished by regularizing a student network to mimic the probabilistic ''softMax" outputs of a teacher network\cite{hinton2015distilling,bucilua2006model} or the intermediate features in a high capacity model\cite{romero2014fitnets, chen2017learning, li2017mimicking,gupta2016cross}. Knowledge distillation has been successfully applied to object detection \cite{chen2017learning}, natural image segmentation \cite{gupta2016cross}, and more recently for medical image analysis\cite{wang2019,murugesan2020,Kats2019}. Model compression is typically meaningful when the high-capacity teacher network is computationally infeasible for real-time analysis\cite{bucilua2006model}. \textcolor{black}{In medical imaging, knowledge distillation using the standard network compression idea has been used for lesion segmentation \cite{wang2019,Kats2019}} using the same modality. However, a key requirement of knowledge distillation methods is the availability of a high-capacity teacher network, pre-trained on a large training corpus. 
	\subsection{Distillation learning as knowledge augmentation}
	A different distillation learning approach considers the problem of increasing knowledge without requiring a large pre-trained teacher network. The problem is then cast as collaborative learning\cite{song2018collaborative,zhang2018CVPR} where multiple weak learners solving the same task are trained collaboratively to improve robustness. Knowledge is added because the networks use different parameter initialization and extract slightly different representations. The key idea here is that increasing robustness improves accuracy. This idea has been shown to be highly effective in self-distillation tasks, where the knowledge learned by a teacher can be refined and improved through hidden self-training of ''seeded" student networks with the same architectural complexity as teacher for classification tasks\cite{FurlanelloICML2018}. The computational complexity of sequential training of learners was recently addressed by using models extracted at previous iterations (as teachers) to regularize the models computed in subsequent iterations (as students)\cite{YangCVPR2019}. However, robustness is defined in the context of achieving consistent inference regardless of the initialization conditions. Although important, improving robustness to initial conditions does not guarantee robustness to imaging conditions. 
	\par
	Knowledge augmentation has also been studied in the context of leveraging different sources of information as additional datasets with additional training regularization\cite{dou2020,Li_Yu_Wang_Heng_2020,KangMICCAI2020,HeDualLearningNeurIPS2016}. Regularization is accomplished by requiring the student network to mimic teacher network's output\cite{KangMICCAI2020,Li_Yu_Wang_Heng_2020}, aligning the feature distribution of the teacher and student modality using a shared network\cite{dou2020}, as well as through cyclically consistent outputs\cite{HeDualLearningNeurIPS2016} of the two networks. 
	\par
	Different from afore-mentioned works, we interpret knowledge augmentation as an approach where the teacher modality (e.g. MRI) is used to guide the extraction of task relevant features from a less informative student modality (e.g CT). 
	\subsection{Medical Image segmentation} 
	\textcolor{black}{Medical image segmentation using deep learning is a well researched topic, with several new architectures\cite{li2018h,zhou2018unet++,jiang2019,dolz2018hyperdense,yu2017automatic} developed for organs\cite{dolz2018hyperdense,oktay2018attention,zhou2018unet++,nikolov2018deep,yu2018recurrent,FuMedPhys2020,Li_Yu_Wang_Heng_2020,dou2020} and tumor segmentation\cite{li2018h,zhao2018tumor,jiang2018tumor,pang2019fast}. Prior works have addressed the issue of low soft tissue contrast on CT by identifying spatially congruent regions by using attention gates\cite{oktay2018attention}, combining multiple views with attention to segment small organs\cite{yu2018recurrent}, as well as squeeze-excite mechanisms to extract relevant features\cite{hu2018squeeze}. Alternatively, computing a highly non-linear representation by combining features extracted at different levels using dense and residual connections have shown to be useful for brain tissues\cite{dolz2018hyperdense} and lung tumor segmentation\cite{jiang2019}. A highly nested formulation of the Unet called Unet++\cite{zhou2018unet++} produced promising accuracies for a large variety of tasks on diverse imaging modalities. Cross-modality distillation learning uses a different perspective, wherein the feature extraction uses explicit guidance from a teacher network \textcolor{black}{to extract features that signal the differences} between the various structures in the image. \textcolor{black}{This approach has demonstrated feasibility for} both natural image\cite{gupta2016cross} and medical image segmentation\cite{dou2020,Li_Yu_Wang_Heng_2020}.} 

	\subsection{Medical Image synthesis for segmentation}
	\textcolor{black}{I2I synthesis has often been used for cross-modality data augmentation \cite{jiang2018tumor,huo2018synseg,Zhang_2018_CVPR,cai2018towards}, with promising accuracies using both semi-supervised\cite{Zhang_2018_CVPR, jiang2018tumor} and unsupervised segmentation\cite{jiang2020TMI,you2020unsupervised,he2021cross} learning settings. I2I synthesis has also been used to compute a different image representation \cite{leiPMB2020,yang2020synthetic} as well as noise reduction on CT\cite{tang2018MLMI} for improved segmentation.}

	\section{Methods}
	\subsection{Cross modality educed distillation (CMEDL)}
	\label{ssec:CMEDL}
\begin{figure}[t]
		\begin{center}
			\includegraphics[width=0.7\columnwidth,scale=0.5]{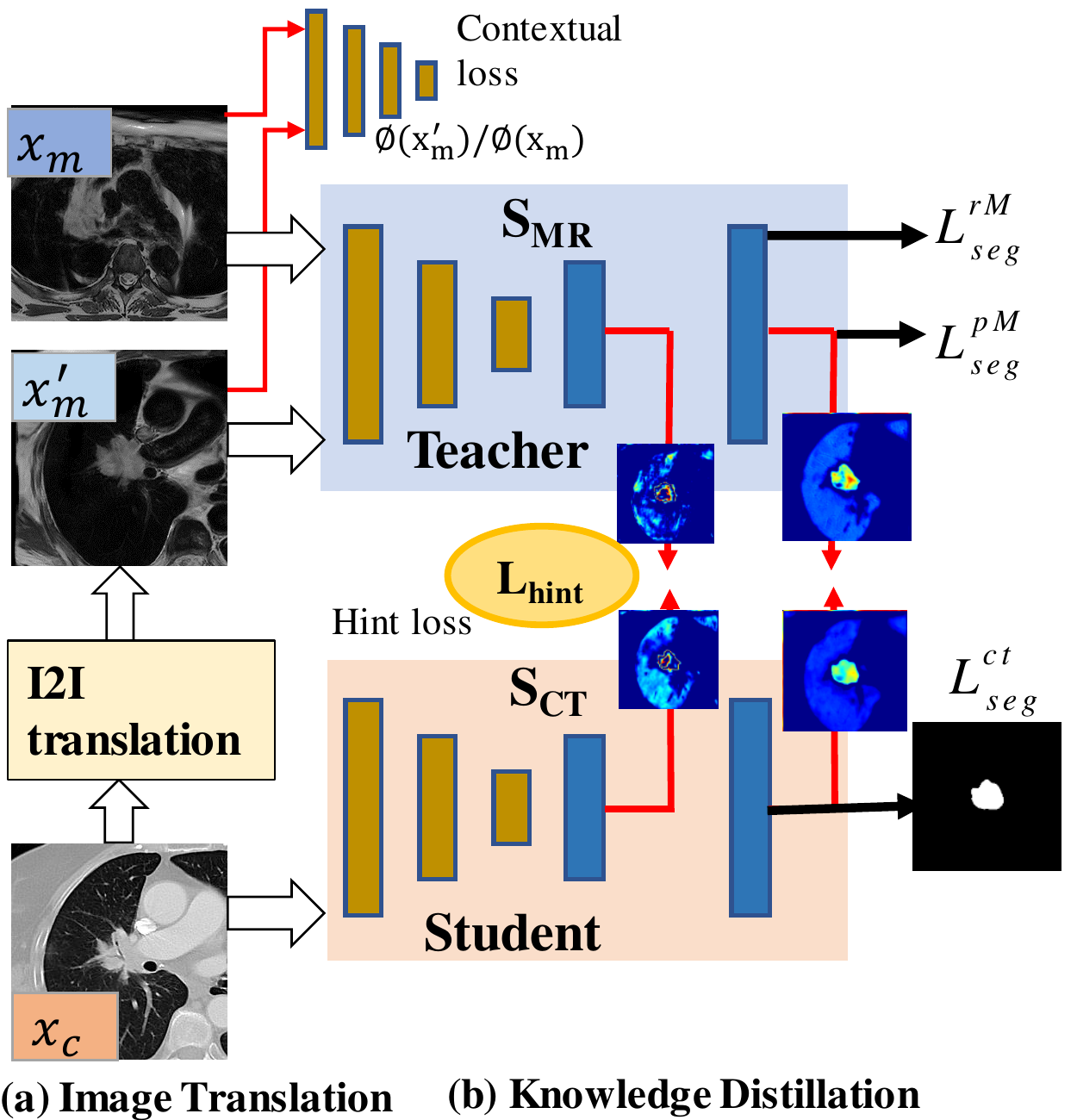}
			\vspace{-0.05cm}
			\setlength{\belowcaptionskip}{-0.4cm}
			\setlength{\abovecaptionskip}{0.08cm}
			\caption{\label{fig:methods} \small Approach overview. $x_{c}$, $x_{m}$ are the CT and MR images from unrelated patient sets; $x_{m}^{'}$ is the pseudo MR image; $S_{MR}$ and $S_{CT}$ are the teacher and student networks, respectively; Network training is optimized with contextual, hint, generator/discriminator losses, and segmentation ($L^{rM}_{seg}$, $L^{pM}_{seg}$, $L^{ct}_{seg}$) losses.}
		\end{center}
	\end{figure}

	An overview of our approach is shown in Fig. \ref{fig:methods}, which consists of cross-modality I2I translation (i.e. CT to pMRI) (Fig.~\ref{fig:methods}a) and knowledge distillation-based segmentation(Fig.~\ref{fig:methods}b) sub-networks for MRI ($S_{MRI}$) and CT ($S_{CT}$). All networks are simultaneously optimized to regularize both pMRI generation, CT and MRI image segmentation. The default I2I network, which uses cycleGAN\cite{zhu2017unpaired} consists of two generators ($G_{C \rightarrow M}$ and $G_{M \rightarrow C}$) for CT to MRI and MRI to CT translation, respectively, two discriminators ($D_{C}$ and $D_{M}$) and a pre-trained VGG19 \cite{simonyan2014very} for calculating the contextual loss \cite{mechrez2018contextual}. 
	
	\subsubsection{Knowledge distillation segmentor}
	Two separate MRI ($S_{MR}$) and CT ($S_{CT}$) segmentation networks with the same architecture to simplify hint loss computation are trained in parallel. The MRI network is trained with both expert segmented T2w MRI ($\{x_{m}, y_{m}\} \in \{X_{M},Y_{M}\}$) and synthesized pseudo MRI (pMRI) ($\{x_{c}^{m}, y_{c}^{m} \}$). The CT network is trained with only CT examples ($\{x_{c}, y_{c}\} \in \{X_{C},Y_{C}\}$). Dice loss is used to optimize networks' training. The loss computed using real MRI data for the MRI network is expressed as $L_{seg}^{rM}$, the loss computed for MRI network using pMRI data is expressed as $L_{seg}^{pM}$, and the loss for the CT network is expressed as $L_{seg}^{CT}$. The total segmentation loss is computed as:
	
	\begin{equation}
	\setlength{\abovedisplayskip}{1pt}
	\setlength{\belowdisplayskip}{1pt}
	\begin{split}
	L_{seg} & = L_{seg}^{rM}+L_{seg}^{pM}+L_{seg}^{CT}\\ &
	=\underset{x_m,x_c}{\mathbb{E}}[-log P(y_{m}|S_{MR}(x_{m}))\\ & -log P(y_{c}|S_{MR}(G_{CT\rightarrow MR}(x_{c})) -log P(y_{c}|S_{CT}(x_{c}))].
	\label{eqn:Seg}
	\end{split}
	\end{equation}
	\par
	The pMRI are used to extract features from MR\textcolor{black}{I} network to compute hint losses for the CT network and also provide additional data to optimize MR\textcolor{black}{I} network using $L_{seg}^{pM}$. The features closest to the output have been shown to be the most correlated to the output task \cite{lin2017refinenet}. Hence, hint loss was computed by minimizing the Frobenius norm of the features from the last two layers of $S_{MRI}$ and $S_{CT}$ networks:
	\begin{equation}
	\setlength{\abovedisplayskip}{1pt}
	\setlength{\belowdisplayskip}{1pt}
	\begin{split}
	L_{hint} & =  \sum_{i=1}^{N} \|\phi{_{CT}^{i}}(x_{c})-\phi{_{MR}^{i}}(G_{CT\rightarrow MR}(x_{c}))||^{2}_{F}, 
	\label{eqn:Feature}
	\end{split}
	\end{equation}
	where $\phi_{CT}^{i}, \phi_{MR}^{i}$ are the $i_{th}$ layer features computed from the two networks, $N$ is the total number of features. 
	
	\subsubsection{Cross modality I2I translation for unpaired distillation}
	This network produces pseudo MRI (pMRI) images. Any cross-modality I2I translation method can be used for this purpose. We demonstrate feasibility with two different methods.
	\paragraph{CycleGAN-based I2I translation} Our default implementation uses a modified cycleGAN\cite{zhu2017unpaired} with contextual losses\cite{mechrez2018contextual} added to better preserve spatial fidelity of structures when using unpaired image sets for training. Contextual loss is implemented by treating an image as a collection of features, where the difference between two images are computed using all-pair feature similarities, which ignores spatial location of features. The contextual image similarity is computed by marginalizing over all the source ($f(G(X_{CT}))$ = ${g_{j}}$) and target image features ($f(X_{MR})$=${m_{i}}$) similarities as:
	\begin{equation}
	CX(g,m) = \frac{1}{N}\sum_{j} \underset{i}max CX(g_{j},m_{i}),
	\end{equation} 
	where, $N$ corresponds to the number of features. The contextual loss is then computed by normalizing the inverse of cosine distances between the features in the two images as:
	\begin{equation}
	L_{cx} = -log(CX(f(G(X_{CT})), f(X_{MR})).
	\end{equation}
    I2I network training is further stabilized using standard adversarial losses ($L_{adv} = L_{adv}^{CT}+L_{adv}^{MR}$), which  maximize the likelihood that the synthesized images (pCT, pMRI) will resemble $X_{CT}$ and $X_{MRI}$.
	\begin{equation}
	\begin{split}
	\setlength{\abovedisplayskip}{0pt}
	\setlength{\belowdisplayskip}{0pt}
	& L^{MRI}_{adv}(G_{C\rightarrow M}, D_{M}, X_{M}, X_{C})= \underset{x_c \sim x_m}{\mathbb{E}}   [log(D_{M}(x_{m})) \\  & + log(1-(D_{M}(G_{C \rightarrow M}(x_{c})) \\
	& L^{CT}_{adv}(G_{M \rightarrow C}, D_{C}, X_{C}, X_{M})  = \underset{x_c \sim x_m}{\mathbb{E}} [log(D_{C}(x_{c})) \\ + & log(1-(D_{C}(G_{M \rightarrow C}(x_{m}))] \\
	\end{split} 
	\label{eqn:adversary loss_MRI}
	\end{equation}
	
	In order to handle translation using unpaired image sets, cycle consistency loss ($L_{cyc}$) \cite{zhu2017unpaired} is computed by minimizing the pixel-to-pixel differences (through L1-norm), between the generated image passing through two GANs (e.g. $ G_{C\circlearrowleft M} = G_{M \rightarrow C}(G_{C\rightarrow M}(x_{c}))$) and the original image (e.g. CT):
	\begin{equation}
	\begin{split}
	& L_{cyc}(G_{C \rightarrow M}, G_{M \rightarrow C},X_{C}, X_{M}) \\ &  =  \underset{x_c \sim x_m}{\mathbb{E}}\left[\left\|G_{C\circlearrowleft M}(x_{c}) - x_{c}\right\|_{1} + 
	 \left\|G_{M\circlearrowleft C}(x_{m}) - x_{m}\right\|_{1}\right].
	\end{split}
	\end{equation}
    The total loss is then computed as: 
    \begin{equation}
	\begin{split}
	& L_{total}^{cyc}= L_{adv} + \lambda_{cyc}L_{cyc} + \lambda_{cx}L_{cx} + \lambda_{hint}L_{hint} + \lambda_{seg}L_{seg}
	\end{split}
	\end{equation}
    where  $\lambda_{cyc}$, $\lambda_{cx}$, $\lambda_{hint}$ and $\lambda_{seg} $ are the weighting coefficients for each loss.
	
	\paragraph{VAE based I2I translation}
	As an alternative I2I translation approach, we implemented a VAE using the Diverse image-to-image translation (DRIT)\cite{lee2018diverse} method. DRIT disentangles the image into domain independent content code ${E_{c}}:{x_c,x_m} \rightarrow  c$  and domain specific style code $E_{s}^{c}: x_{c} \rightarrow s_{c}$ for $x_{c} \in X_{C}$ and $E_{s}^{m}: x_{m} \rightarrow s_{m}$ for $x_{m} \in X_{M}$. $E_{c}$ is the content encoder while $E_{s}^{c}$ and $E_{s}^{m}$ are the domain specific style encoders corresponding to rendering in the CT and MR domains, respectively. Content adversarial loss to used optimize the domain content encoding:

\begin{equation}
    \setlength{\abovedisplayskip}{1pt}
    \setlength{\belowdisplayskip}{1pt} 
   \begin{split}
      L_{adv}^{c} & = \underset{x_c \sim x_m}{\mathbb{E}}[ log(D_c(E_c(x_c)))+ (1-log(D_c(E_c(x_m)))) \\ + & log(D_c(E_m(x_m)))+(1-log(D_c(E_m(x_c))))]
  \end{split}
\end{equation}

\begin{figure*}
		\begin{center}
			\includegraphics[width=1.6\columnwidth,scale=1]{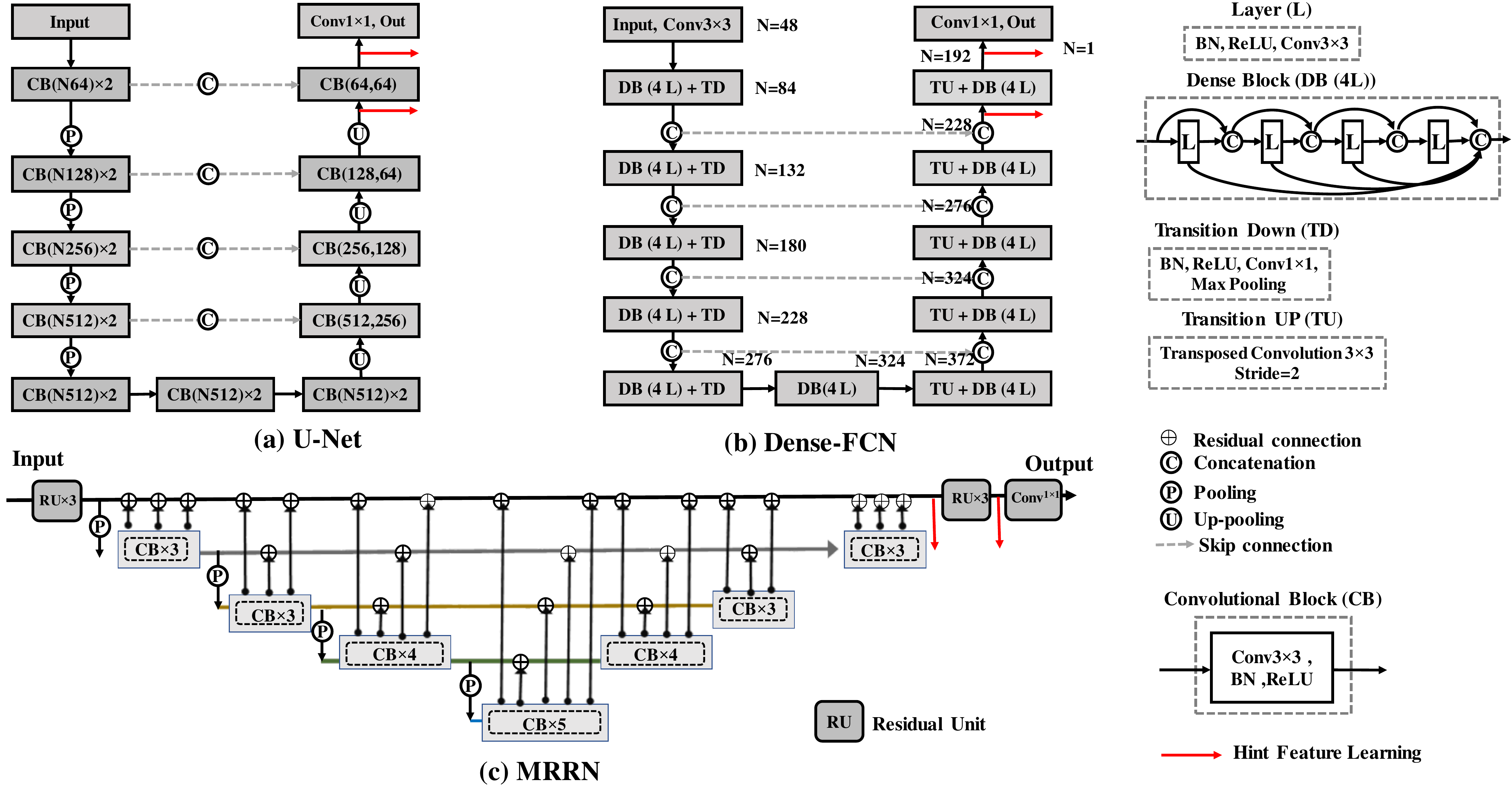}
			\vspace{-0.05cm}\setlength{\belowcaptionskip}{-0.4cm}\setlength{\abovecaptionskip}{0.08cm}\caption{\label{fig:seg_structure} \small The segmentation structure of Unet \cite{ronneberger2015u} and DenseFCN57 \cite{jegou2017one}. The red arrow indicates that the output of these layers are used for distilling information from MR into CT. This is done by minimizing the L2-norm between the features in these layers between the two networks. The blue blocks indicate the lower layer; the green blocks indicate the middle layer; the orange blocks indicate the upper layer in Unet.}
		\end{center}
	\end{figure*} 

We compute a content code reconstruction loss $L_{cc}$:
$\hat{x}_{j} =  G(E_{c}(x_i), E_{s}(x_j), d_j)$, where $i$ is the source domain and $j$ is the translated target domain. 
\begin{equation}
\setlength{\abovedisplayskip}{1pt}
\setlength{\belowdisplayskip}{1pt} 
\begin{split}
L_{cc} = E \| E_{c}(x_{c}) - E_{c}(\hat{x}_{c})\|_{1} + E \| E_{m}(x_{m}) - E_{m}(\hat{x}_{m})\|_{1}. 
\end{split}
\end{equation}
$\hat{x_c}, \hat{x_m}$ are computed as $\hat{x_c}=G_c(E_c(x_c),E_s^m(x_m))$ and $\hat{x_m}=G_m(E_m(x_m),E_s^c(x_c))$, respectively.
The domain specific style encodings $E_{s}^{c}, E_{s}^{m}$ are extracted by minimizing the KL-divergence of a latent encoding computed using a conditional VAE with respect to the corresponding image domains:
\begin{equation}
\setlength{\abovedisplayskip}{1pt}
\setlength{\belowdisplayskip}{1pt} 
\begin{split}
L_{VAE} = & \underset{x_c \sim X_C}{\mathbb{E}}[D_{KL}(E_s^c(x_c)||q_c(x_c) )] + \|\hat{x_c} - x_c \|_{1} \\ & + 
 \underset{x_m \sim X_M}{\mathbb{E}}[D_{KL}(E_s^m(x_m)||q_m(x_m))] + \|\hat{x_m} - x_m \|_{1},
 	\end{split}
\end{equation}  
$q_c(x_c)$ and $q_m(x_m)$ are prior normal distributions with unit covariance $\mathcal{N}$(0,$I$), and  $\hat{x_c}$=$G_c(E_c(x_c),E_s^m(x_m))$ and $\hat{x_m}$=$G_m(E_m(x_m),E_s^c(x_c))$, respectively. Adversarial losses are used to optimize image generation:
\begin{equation}
\setlength{\abovedisplayskip}{1pt}
\setlength{\belowdisplayskip}{1pt} 
\begin{split}
  L_{adv} = & \underset{x_c\sim X_{C},x_m \sim X_{M}}{\mathbb{E}} [log(D(x_m))+ \\
  & 0.5 \times log(1-D(G_c(E_c(x_c),E_m(x_m)))]+ \\
  & \underset{z \sim N(0,1)}{\mathbb{E}}[0.5\times log(1-(D(G_c(E_c(x_c),z))]
 \end{split}
\end{equation}
where z is sampled from $\mathcal{N}$(0,$I$). In addition, latent code regression loss $L_{lr}$ is used to regularize I2I translations. The latent regression loss is computed as:

\begin{equation}
\setlength{\abovedisplayskip}{1pt}
\setlength{\belowdisplayskip}{1pt} 
\begin{split}
L_{lr} = &\underset{z \sim N(0,1)}{\mathbb{E}} \| z - E_{s}^c(G_c(E_{c}(x_c), z))\|_{1} + \\ &\underset{z \sim N(0,1)}{\mathbb{E}} \| z - E_{s}^m(G_m(E_{m}(x_m), z))\|_{1}. 
\end{split}
\end{equation}

The VAE loss is computed as:
\begin{equation}
	\begin{split}
	L_{total}^{DRIT}= & L_{adv}+ \lambda_{c}{L_{adv}^c}+ \lambda_{vae}{L_{VAE}}+ \lambda_{lr}L_{lr}+ \lambda_{cc}{L_{cc}}+ \\ & \lambda_{hint}L_{hint}+ \lambda_{seg}L_{seg}
	\end{split}
\end{equation}

where  $\lambda_{c}$, $\lambda_{vae}$, $\lambda_{lr}$, $\lambda_{cc}$, $\lambda_{hint}$ and $\lambda_{seg} $ are the weighting coefficients for each loss.
    
\subsubsection{Optimization}
Teacher and student segmentors, I2I translation generators and discriminators are trained jointly and end to end. Network update alternates between the I2I translation for teacher modality (e.g. pMRI) generation and knowledge distillation based student modality (e.g. CT) segmentation with the following gradients, $-\Delta_{\theta_{G}}(L_{adv})$+ $\lambda_{cyc}${$L_{cyc}$}+ $\lambda_{CX}{L_{CX}}$+ $\lambda_{hint}L_{hint}$+ $\lambda_{seg}L_{seg}$, -$\Delta_{\theta_{D}}$ $(L_{adv})$ and $-\Delta_{\theta_{S}}(L_{hint}$+$L_{seg})$. 
\par
The VAE gradients are computed as, $-\Delta_{\theta_{G}}(L_{adv}+ \lambda_{c}{L_{adv}^c}+ \lambda_{vae}{L_{VAE}}+ \lambda_{lr}L_{lr}+ \lambda_{cc}{L_{cc}}+ \lambda_{hint}L_{hint}+ \lambda_{seg}L_{seg})$, $-\Delta_{\theta_{D}}(L_{adv})$ and $-\Delta_{\theta_{S}}(L_{hint}+L_{seg})$.

\subsection{Networks architecture details:}
	
\paragraph{I2I translation network}
Details of cycleGAN and VGG16 network used for computing contextual loss are in our prior work\cite{jiang2019MICCAI}. Briefly, generators were implemented using two stride 2-convolutions, 9 residual blocks, and fractionally strided convolutions with half strides, and discriminators using 70$\times$70 patchGAN. Contextual loss was computed by extracting the higher level features using (after Conv7, Conv8, and Conv9 with a feature size of 64$\times$64$\times$256, 64$\times$64$\times$256 and 32$\times$32$\times$512) from a pre-trained VGG16 (trained on the ImageNet database) to accommodate limited GPU memory.

The VAE network was based on the DRIT\cite{lee2018diverse} method. The content encoder $E_c$ was implemented using a fully convolutional network and the style encoders $E_{s}^{c}, E_{s}^{m}$ were composed of several residual blocks followed by global pooing and fully connected layers, with the output layer implemented using a reparameterization trick. Generator networks used 6 residual blocks. 

\paragraph{Segmentation networks structure}
	We implemented the Unet \cite{ronneberger2015u}, DenseFCN \cite{jegou2017one}, and multiple resolution residual network (MRRN)\cite{jiang2019} segmentation methods. The Unet and DenseFCN architectures are described in more detail in our prior work\cite{jiang2019MICCAI}. 
	
	\textbf{The Unet }\rm network used 4 max-pooling and 4 up-pooling layers with skip connections to concatenate the low-level and high-level features. Batch normalization (BN) and \textit{ReLU} activation were used after the convolutional blocks. Feature distillation was done using the last two layers feature size of 128$\times$128$\times$64 and 256$\times$256$\times$64 are used to tie the features, shown as red arrow in Fig.\ref{fig:seg_structure} (a). This network had 13.39 M parameters and 33 layers\footnote{layers are only counted on layers that have tunable weights}. 
	
	\textbf{The DenseFCN}\rm network used dense blocks with 4 layers for feature concatenation, 5 transition down for feature down-sampling, and 5 transition up blocks for feature up-sampling with a growing rate of 12. Hint losses were computed using features from the last two blocks of DenseFCN with feature size of 128$\times$128$\times$228 and 256$\times$256$\times$192, shown as red arrow in Fig.\ref{fig:seg_structure} (b). This network had 1.37 M parameters and 106 layers.

    \textbf{Multiple resolution residual network (MRRN):}\rm The MRRN\cite{jiang2019} is a very deep network that we previously developed for lung tumor segmentation. This network incorporates aspects of both densely connected\cite{huang2017densely} and residual networks\cite{HeResidualNet2016} by combining features computed at multiple image resolutions and layers. Feature combination is done using residual connection units (RCU). RCU takes two inputs, feature map from the immediately preceding network layer or the output of preceding RCU and the feature map from the residual feature stream. A new residual stream is generated following each downsampling operation in the encoder. The residual feature streams thus carry feature maps at specific image resolutions for combination with the deeper layer features. Four max-pooling and up-pooling are used in the encoder and decoder in MRRN. Feature distillation was implemented using features from the last two layers of size 128$\times$128$\times$128 and 256$\times$256$\times$64, as shown in by the red arrow in Fig.~\ref{fig:seg_structure}(c).
    This network had 38.92M parameters.

\subsection{Implementation }
All networks were implemented using the Pytorch \cite{paszke2017automatic} library and trained end to end on Tesla V100 with 16 GB memory and a batch size of 2. The ADAM algorithm was used for optimization. An initial learning rate of 1e-4 was used for I2I networks and 2e-4 for the segmentation networks. We set $\lambda_{adv}$=1, $\lambda_{cyc}$=10, $\lambda_{CX}$=1, $\lambda_{hint}$=1 and $\lambda_{seg}$=5 for training CMEDL with CycleGAN. We set $L_{c}$=1, $\lambda_{vae}$=1, $\lambda_{cc}$=10, $\lambda_{lr}$=10, $\lambda_{hint}$=1 and $\lambda_{seg}$=5 for training VAE-CMEDL. 
\textcolor{black}{Hyperparameters were used as is from CycleGAN and VAE-DRIT networks. $\lambda_{seg}$=5 was used as in\cite{jiang2018tumor,jiang2020TMI}. Parameter $\lambda_{hint}$=1 was determined empirically using the default CMEDL network as described in the Supplementary document Sec. IV.} 
\\
\textcolor{white}{AA}Online data augmentation using horizontal flip, scaling, rotation, elastic deformation were applied to ensure generalizable training with sufficient data. The segmentation validation loss was monitored during the training to prevent over-fitting through early stopping strategy with a maximum training epoch of 100. We will make our code available through GitHub upon acceptance for publication.

\section{Experiments and Results}
\subsection{Evaluation metrics}
Segmentation accuracies were measured using the Dice similarity coefficient (DSC), contour surface distance (SDSC)\cite{nikolov2018deep}, and Hausdroff distance (95\%) or HD95. \textcolor{black}{pMRI synthesis accuracy was computed using Kullback–Leibler (KL) divergence\cite{jiang2018tumor}, peak signal to noise ratio (PSNR), and structural similarity index (SSIM) measures. Details of the accuracy metrics are in Supplementary document. Statistical comparisons to establish segmentation accuracy differences were computed using two-sided paired Wilcoxon signed rank tests of the analyzed and CMEDL methods at 95$\%$ significance level.}
 
\subsection{Datasets}
\subsubsection{Cross-modality (CT-MR\textcolor{black}{I}) distillation for tumor segmentation}
\textbf{CT lung tumor dataset: \/}\rm CT scans of patients diagnosed with locally advanced non-small cell lung cancer (LA-NSCLC) and treated with intensity modulated radiation therapy (IMRT) and sourced from both internal archive and open-source NSCLC-TCIA dataset\cite{aerts2015data} were analyzed. Training used 377 cases from the NSCLC-TCIA; validation (N = 209 tumors from 50 patients), and testing (N = 609 tumors from 177 patients) used internal archive patients. Both external and subset of the internal datasets were used in our prior work\cite{jiang2019}. Segmentation robustness was computed with respect to five radiation oncologists using twenty additional cases from an open-source lung tumor dataset in patients treated with radiation therapy\cite{kalendralis2020fair}. Networks were trained with 58,563 2D CT image patches and 42,740 MR\textcolor{black}{I} image patches of size 256$\times$256 pixels enclosing the tumor and the chestwall. \textcolor{black}{All the image slices containing the lungs were used for testing. This was accomplished by automatically identifying the lung slices by intensity threshold (HU \textless -300), followed by connected regions extraction to extract the largest 2-component that 
indicate the left and right lungs.}
\\
\textcolor{white}{AA}\textbf{MRI lung tumor dataset: \/}\rm Eighty one T2w turbo spin echo MRI images acquired weekly from 28 LA-NSCLC patients treated with definitive IMRT at our institution, as described in our prior work\cite{jiang2018tumor} was used. 
\subsubsection{Same modality distillation for MRI multi-organ segmentation} 
\textcolor{black}{We used 20 T1-DUAL in-phase MRI (T1w) and T2w spectral pre-saturation inversion recovery MRI from the ISBI grand challenge Combined Healthy Abdominal Organ Segmentation (CHAOS) challenge data\cite{CHAOS2019}. Segmented organs included the liver, spleen, left and right kidney. Histogram standardization, MRI signal intensity clipping (T1w in range 0 to 1136; T2w in range 0 to 1814), followed by 2D patch extraction (256$\times$256 pixels) was done. \textcolor{black}{Three-fold cross-validation using 8000 T1w and 7872 T2w MRI image patches, taking care patient slices did not fall into different folds was done. Results from the validation folds not used in training are reported.}}
\\
\textcolor{white}{AA}\textcolor{black}{More details of the CT/MR image protocols for all datasets are in the Supplementary document Sec. I.}

\subsection{Experiments}
\subsubsection{Impact of segmentation and I2I architectures on accuracy}
\begin{figure}[h]
				\begin{center}
					\includegraphics[width=1.0\columnwidth,scale=1]{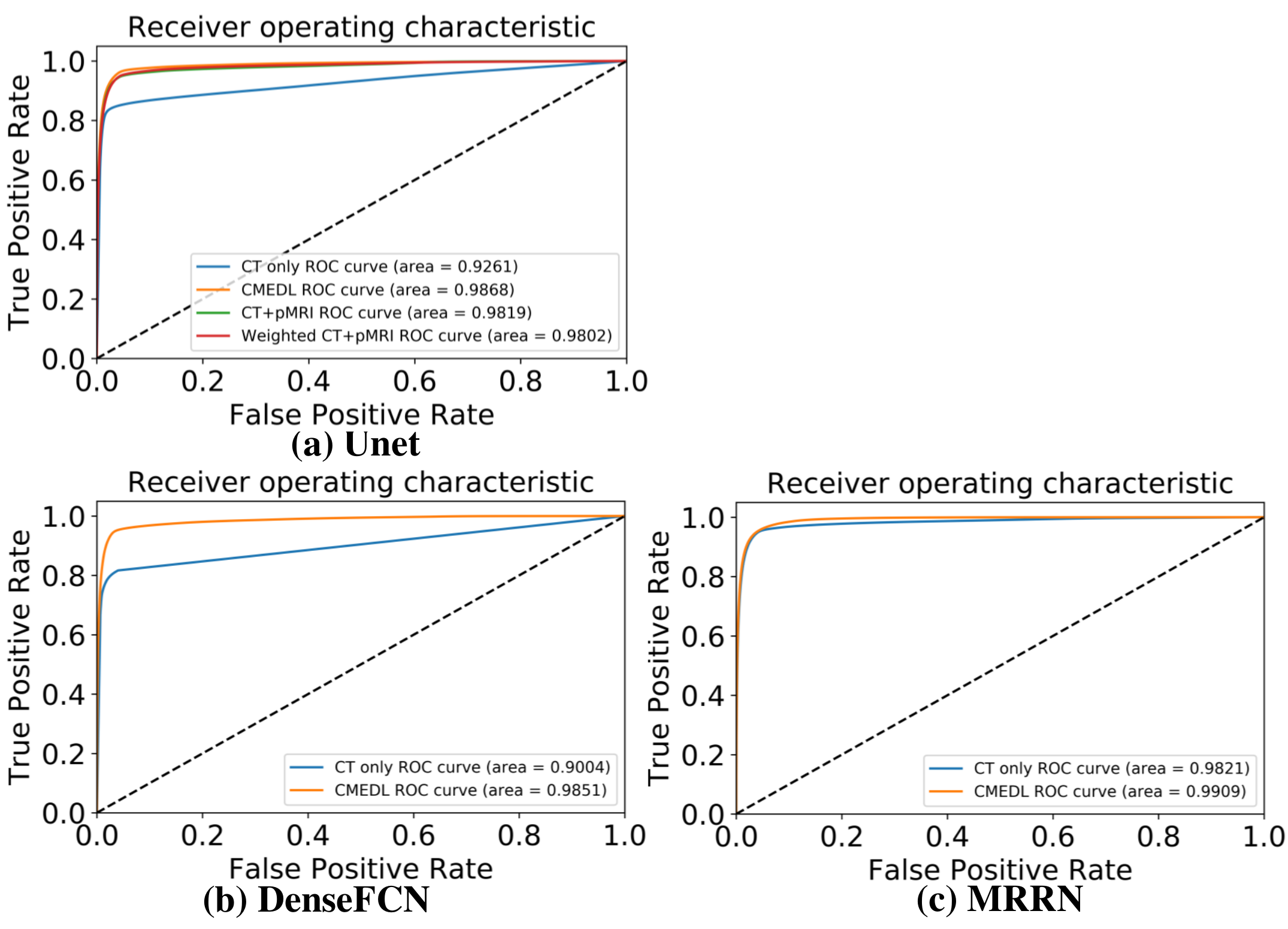}
					\vspace{-0.05cm}\setlength{\belowcaptionskip}{-0.4cm}\setlength{\abovecaptionskip}{0.08cm}\caption{\label{fig:roc_curve} \small \textcolor{black}{The ROC curves of different network implementations. ROC curves computed with different CT and MR fusion strategies using the Unet is also shown.}}
				\end{center}
			\end{figure}

Table.~\ref{tab:segmentationAcc_val_test_tumor_Unet} shows the segmentation accuracy on test sets computed for the various segmentation architectures (Unet, denseFCN, and MRRN) with and without the CMEDL approach. Also, accuracies when using cycleGAN with contextual loss corresponding to the standard CMEDL and with a DRIT VAE network (VAE-CMEDL). \textcolor{black}{Significantly accurate results are indicated (Table.~\ref{tab:segmentationAcc_val_test_tumor_Unet}) with an asterisk.} The CMEDL approach was significantly more accurate than the non-CMEDL methods (MRRN p $<$ 0.001; Unet p $<$ 0.001; dense-FCN p $<$ 0.001) for all accuracy metrics, \textcolor{black}{while requiring the same computational resources as the non-CMEDL methods for testing (Unet with 8.1ms, DenseFCN with 11.7ms, and MRRN with 17.8ms)\footnote{\textcolor{black}{Testing time is calculated as the inference time for one single image with size of $256$ $\times$ $256$ on Nvidia V100 GPU.} }. Details of networks' parameters, training, and testing times are in Supplementary Table I. Both standard CMEDL and VAE-CMEDL produced similar accuracies (Table \ref{tab:segmentationAcc_val_test_tumor_Unet}), although VAE-CMEDL required longer time for each gradient update during training (Supplementary Table I).} \textcolor{black}{Fig~\ref{fig:roc_curve} shows the receiver operating curves (ROC) for tumor segmentation using the various network implementations. All methods show a clear performance difference between CMEDL and CT only segmentation.}

\begin{table}[ht] 
	\centering{\caption{Segmentation accuracy for lung tumors from CT using Unet, DenseFCN and MRRN on Test Set. \textcolor{black}{$^{*}$ indicates significant difference with p $<$ 0.05.}}
	\label{tab:segmentationAcc_val_test_tumor_Unet}
	\centering
	\scriptsize
	\setlength{\tabcolsep}{1.5mm}
	\begin{tabular}{|c|c|c|c|c|} 
		\hline 
		\multirow{2}{*}{Network}&\multirow{2}{*}{Method}&  \multicolumn{3}{|c|}{Testing (N=609 lung tumors) CT}\\ 
		\cline{3-5}
		\cline{3-5}
		{}&{    }  & {  DSC  ($\uparrow$)}& {  SDSC ($\uparrow$)}& {  HD95 mm ($\downarrow$) }\\
        \hline
        \multirow{3}{*}{Unet}&{CT only} & { 0.69$\pm$0.20\textcolor{black}{$^{*}$}}& {  0.73$\pm$0.21\textcolor{black}{$^{*}$}} & { 13.44$\pm14.69$\textcolor{black}{$^{*}$}}\\
		{}&\multirow{1}{*}{\textcolor{black}{VAE-CMEDL}} & { 0.74$\pm$0.17 }& { 0.79$\pm$0.20 }& { 7.12$\pm$9.28 }\\
		{}&\multirow{1}{*}{CMEDL} & { 0.75$\pm$0.17 }& { 0.81$\pm$0.20 }& { 6.48$\pm$10.33 }\\
		\hline				
        \multirow{3}{*}{DenseFCN}&	\multirow{1}{*}{CT only} & { 0.67$\pm$0.18\textcolor{black}{$^{*}$} }& {0.71$\pm$0.20\textcolor{black}{$^{*}$  }}   & { 13.80$\pm$14.23\textcolor{black}{$^{*}$ }}\\
		
		{}&	\multirow{1}{*}{\textcolor{black}{VAE-CMEDL}} & { 0.73$\pm$0.20 }& { 0.78$\pm$0.23}  & { 7.25$\pm$11.88} \\
	
		{}&	\multirow{1}{*}{CMEDL} & { 0.74$\pm$0.18 }& { 0.79$\pm$0.21}  & { 6.57$\pm$10.29} \\
		\hline
	    \multirow{3}{*}{MRRN}&	\multirow{1}{*}{CT only} & { 0.73$\pm$0.17\textcolor{black}{$^{*}$}}& { 0.78$\pm$0.21\textcolor{black}{$^{*}$ }}& { 6.75$\pm$10.08\textcolor{black}{$^{*}$ }}\\
		{}&	\multirow{1}{*}{\textcolor{black}{VAE-CMEDL}}& { 0.76$\pm0.13$ }& { 0.82$\pm$0.16 }  & { 5.56$\pm7.18$ }\\
	    {}&	\multirow{1}{*}{CMEDL}& { 0.77$\pm0.13$ }& { 0.83$\pm$0.16 }  & { 5.20$\pm6.86$ }\\
		\hline	
	\end{tabular} 
	}
\end{table}

\begin{figure*}
	\begin{center}
		\includegraphics[width=1.90\columnwidth,scale=1]{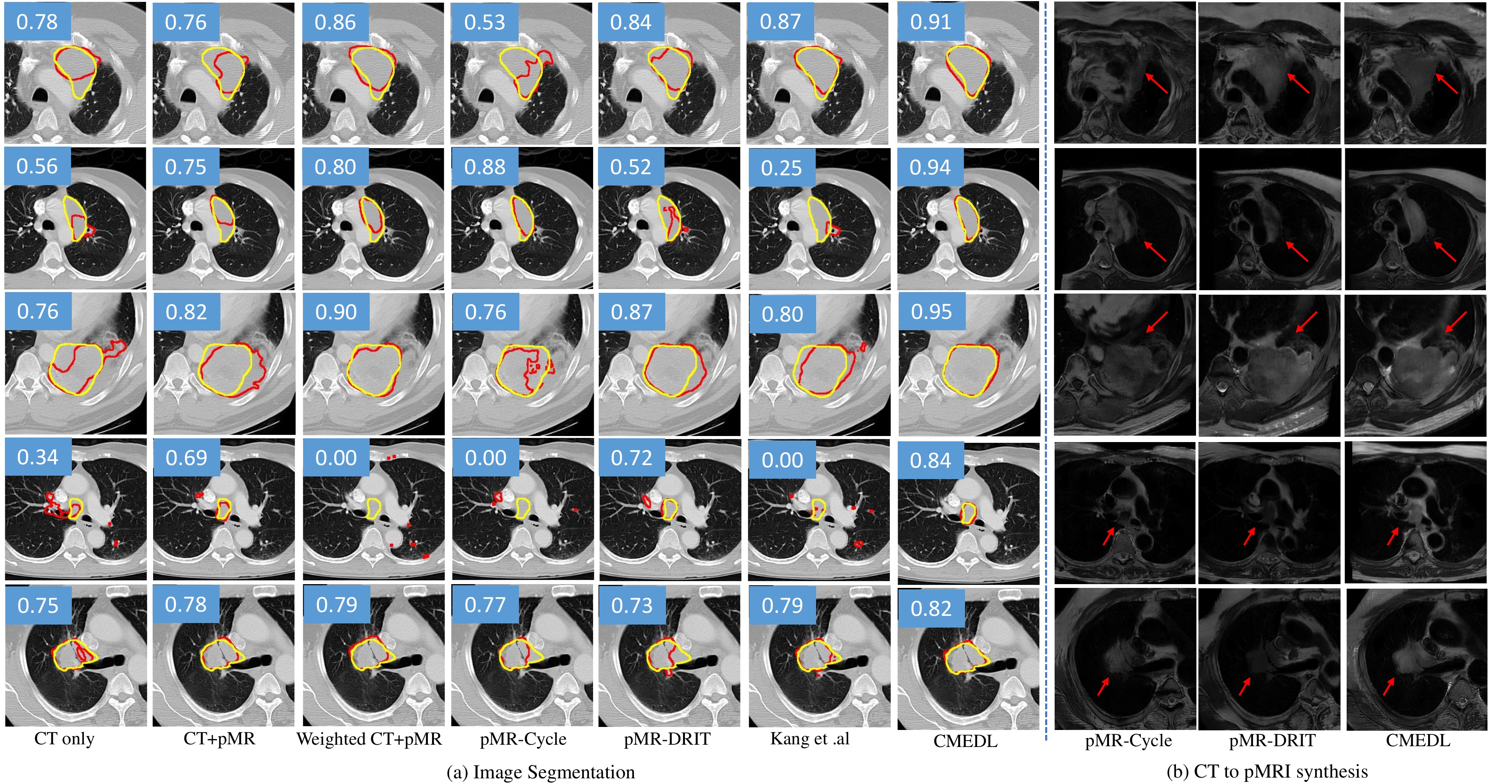}
	\vspace{-0.05cm}\setlength{\belowcaptionskip}{-0.4cm}\setlength{\abovecaptionskip}{0.08cm}\caption{\label{fig:tumors_trans_seg_val_test} \small \textcolor{black}{(a)} Lung tumor segmentations produced by the various methods. Volumetric DSC accuracy is also shown for these methods. Yellow contour corresponds to the expert and red to the algorithm segmentation. \textcolor{black}{(b) shows the pMR images produced by the cycleGAN, DRIT, and CMEDL methods. The tumors are indicated by an arrow on the pMRI.}}
\end{center}
				
\end{figure*}	

\subsubsection{Segmentation accuracy with different pMR\textcolor{black}{I} fusion strategies} 
We evaluated the accuracy when using pMRI with different combination strategies with CT. As voxel-wise fidelity in pMRI synthesis is crucial when using pMRI for segmentation, we evaluated accuracy when using a cycleGAN\cite{zhu2017unpaired}, DRIT-VAE\cite{lee2018diverse}, and the I2I network trained with the standard CMEDL framework.  \textcolor{black}{We also evaluated the accuracy when using the teacher instead of student network to generate segmentations, where the CT image is transformed in to a pMRI image even for testing (pMRI-CMEDL). We also compared our results to a recent cross-domain distillation method \cite{Li_Yu_Wang_Heng_2020}.} Finally, we computed segmentations using using row-wise concatenated pMRI and CT images as input to a segmentation network similar to\cite{FuMedPhys2020} and a weighted-CT+pMRI approach, that weights the relative contribution of pMRI and CT both during training and inference. 
\begin{table} 
		\centering{
			\caption{MR fusion strategies for informative teacher distillation. $^{\dagger}$ refers to network optimized using CMEDL; $^{\star}$ two modalities are channel-wise concatenated.} 
			\label{tab:difference of different methods} 
			\scriptsize
			\begin{tabular}{|c|c|c|c|c|c|} 
				\hline 
				& \multicolumn{2}{c|}{Testing} & \multicolumn{3}{c|}{Training} \\ \hline
				{  Method  }  & {Segmentor}  & { pMR\textcolor{black}{I}} & { Context} & {Hint} & {pCT}\\
				& & synthesis & loss & loss & augment \\
				\hline     
				{   CMEDL }  & {  CT  } & { $\times$ } & { $\checkmark$ }& { $\checkmark$ } & {$\times$}\\ 
				\hline 
				{pMRI$^{\dagger}$} & {MR} & {$\checkmark$} & {$\checkmark$} & {$\checkmark$} & {$\times$}\\ 
				\hline
				{pMRI} & {MR} & {$\checkmark$} & {$\times$} & {$\times$} & {$\times$}\\
				\hline
				{CT+pMRI$^{\dagger}$}  & {CT+pMR$^{\star}$ } & { $\checkmark$ } & { $\times$ }& { $\times$ } & {$\times$}\\
				\hline
				{\cite{KangMICCAI2020}} & {CT} & {$\times$} & {$\times$} & {$\times$} & {$\checkmark$} \\
				\hline
				
			\end{tabular}
		} 
\end{table}

\textbf{Weighted pMRI concatenation: \/}\rm This method accounts for variability in the accuracy of the generated pMR\textcolor{black}{I} images by predicting the relative contribution $\alpha$ of pMRI for CT segmentation. The parameter $\alpha$ was computed using a ResNet18 \cite{HeResidualNet2016} network with 2 fully connected layers, which used the CT and the corresponding pMRI as its inputs. The weighted combination is implemented as:  
\begin{equation}
	\setlength{\abovedisplayskip}{1pt}
	\setlength{\belowdisplayskip}{1pt}
	L_{seg} = \underset{x_c \sim X_{C}}{\mathbb{E}}[-log S((y_{c}|(1-\alpha)x_{c}; \alpha x_{m}^{'}))] 
	\label{eqn:C4}
\end{equation}
Table.~\ref{tab:difference of different methods} shows the differences between the various methods.  

\begin{table}[h] 
	\centering{\caption{CT Unet segmentation accuracy with pMRI fusion strategies. \textcolor{black}{$^{*}$ indicates significant difference.} } 
			\label{tab:connect_stragety} 
			\centering
			\scriptsize
			
			\begin{tabular}{|c|c|c|c|} 
				\hline

				\hline
				{  Method  (U-net)}  & {  DSC ($\uparrow$) }& {  SDSC ($\uparrow$) }& { HD95 mm ($\downarrow$) }\\
				\hline 
				
				\hline
				\multirow{1}{*}{CT only } & { 0.69$\pm0.20$\textcolor{black}{$^{*}$} } & { 0.73$\pm0.21$\textcolor{black}{$^{*}$ }} & { 13.44$\pm14.69$\textcolor{black}{$^{*}$ }}\\
				\multirow{1}{*}{pMRI-Cycle } & { 0.71$\pm0.18$\textcolor{black}{$^{*}$} } & { 0.75$\pm0.20$\textcolor{black}{$^{*}$ }} & { 12.69$\pm13.21$\textcolor{black}{$^{*}$ }}\\
				\multirow{1}{*}{pMRI-\textcolor{black}{DRIT}  }& { 0.71$\pm0.17$\textcolor{black}{$^{*}$} } & { 0.76$\pm0.22$\textcolor{black}{$^{*}$ }} & { 12.47$\pm11.25$\textcolor{black}{$^{*}$ }}\\
				\hline	
				\multirow{1}{*}{ CT+pMR\textcolor{black}{I} } & { 0.72$\pm0.17$\textcolor{black}{$^{*}$} } & { 0.77$\pm0.22$\textcolor{black}{$^{*}$ }} & { 11.50$\pm12.69$\textcolor{black}{$^{*}$ }}\\
				
				\multirow{1}{*}{ Weighted CT+pMR\textcolor{black}{I} } & { 0.72$\pm0.18$\textcolor{black}{$^{*}$} } & { 0.77$\pm0.20$\textcolor{black}{$^{*}$ }} & { 11.40$\pm12.23$\textcolor{black}{$^{*}$ }}\\

				\multirow{1}{*}{pMRI-CMEDL}  & { 0.74$\pm$0.17} & { 0.79$\pm0.20$} & { 8.67$\pm10.45$ }\\
                \hline

				\multirow{1}{*}{\textcolor{black}{Kang et. al. \cite{Li_Yu_Wang_Heng_2020}}}  & { \textcolor{black}{0.72$\pm$0.17\textcolor{black}{$^{*}$}} } & { \textcolor{black}{0.77$\pm$0.23\textcolor{black}{$^{*}$} }} & { \textcolor{black}{12.03$\pm$13.64\textcolor{black}{$^{*}$} }}\\

				\hline					
				\multirow{1}{*}{VAE-CMEDL} & { 0.74$\pm$0.17 } & { 0.79$\pm$0.20 } & { 7.12$\pm$9.28 }\\				
				\multirow{1}{*}{CMEDL} & { 0.75$\pm$0.17 } & { 0.81$\pm$0.20 } & { 6.48$\pm$10.33 }\\
				
				\hline	
				
				\hline
			\end{tabular}
			}
		\end{table}
	
Table.~\ref{tab:connect_stragety} shows the accuracy of CMEDL compared to the multiple fusion strategies, \textcolor{black}{with significant differences indicated by an asterisk.} CMEDL approach was significantly more accurate (p $<$ 0.001) than all pMRI combination methods regardless of the I2I method used. On the other hand, the pMRI-CMEDL network provided similarly accurate segmentations as the default CMEDL method (p $=$ 0.64). \textcolor{black}{However, pMRI-CMEDL requires additional computation due to the need for pMRI synthesis as opposed to the default CMEDL framework, which directly uses the CT segmentor during testing.}
\textcolor{white}{AA}Fig.~\ref{fig:tumors_trans_seg_val_test} (a) shows segmentations produced by various methods on  \textcolor{black}{randomly selected representative test cases}. As shown, the CMEDL method successfully segmented the tumors, even for those tumors with unclear boundaries. \textcolor{black}{CT only method in general performed worse than all other methods (Table.~\ref{tab:connect_stragety}). However, the pMR-DRIT and pMR-Cycle methods produced worse accuracies than even the CT only method (case 2, case 5 for pMRI-DRIT; case 4 for pMRI-Cycle) because of \textcolor{black}{inaccurate pMRI synthesis} (Fig.~\ref{fig:tumors_trans_seg_val_test} (b)) used to generate the segmentation. Simple fusion method (CT+pMRI) was more \textcolor{black}{accurate} than the other pMRI-based segmentation methods, possibly due to inclusion of CT with pMRI for segmentation.} 

\begin{figure*}
	\begin{center}
	    \includegraphics[width=1.8\columnwidth,scale=1]{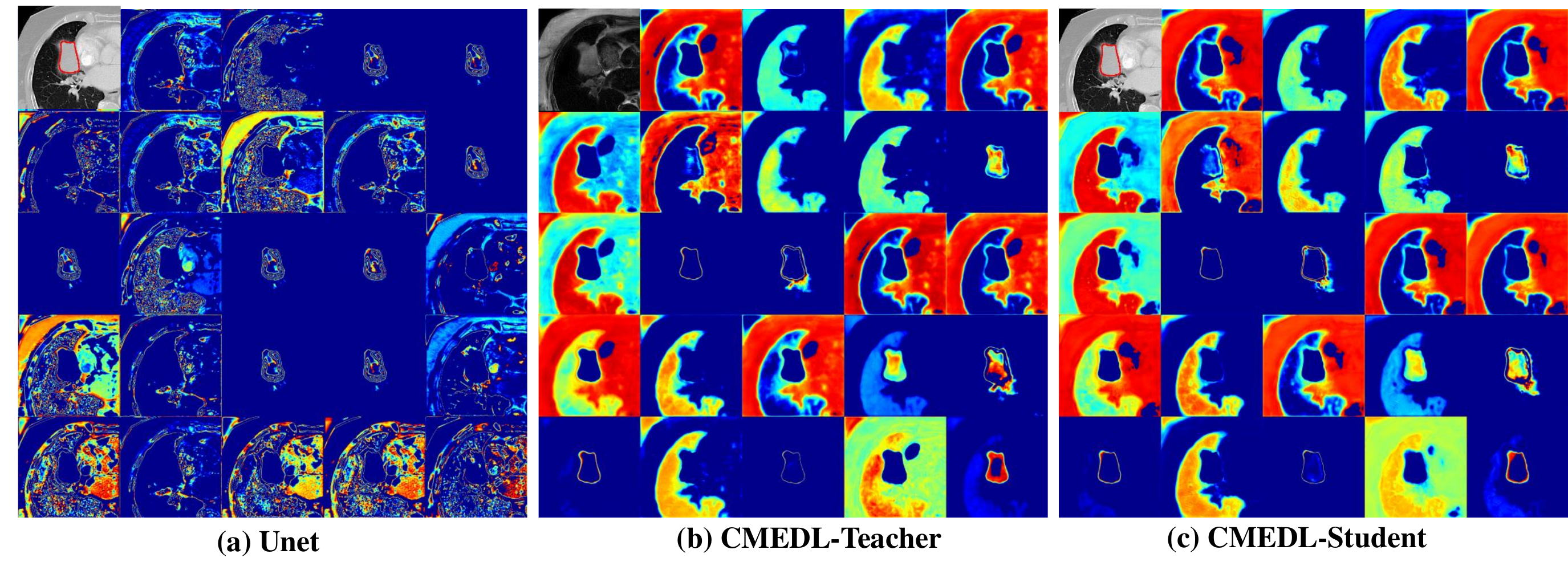}
			\vspace{-0.05cm}\setlength{\belowcaptionskip}{-0.4cm}\setlength{\abovecaptionskip}{0.08cm}\caption{\label{fig:feature_map} \small Feature maps (1 to 24) from last layer of (a) CT only Unet (b) teacher network of CMEDL with synthesized pMRI, and (c) student network of CMEDL. The tumor delineation is also shown on the CT scan.  
		}
	\end{center}
	
\end{figure*}
\begin{figure}[t]
	\begin{center}
		\includegraphics[width=0.80\columnwidth,scale=1]{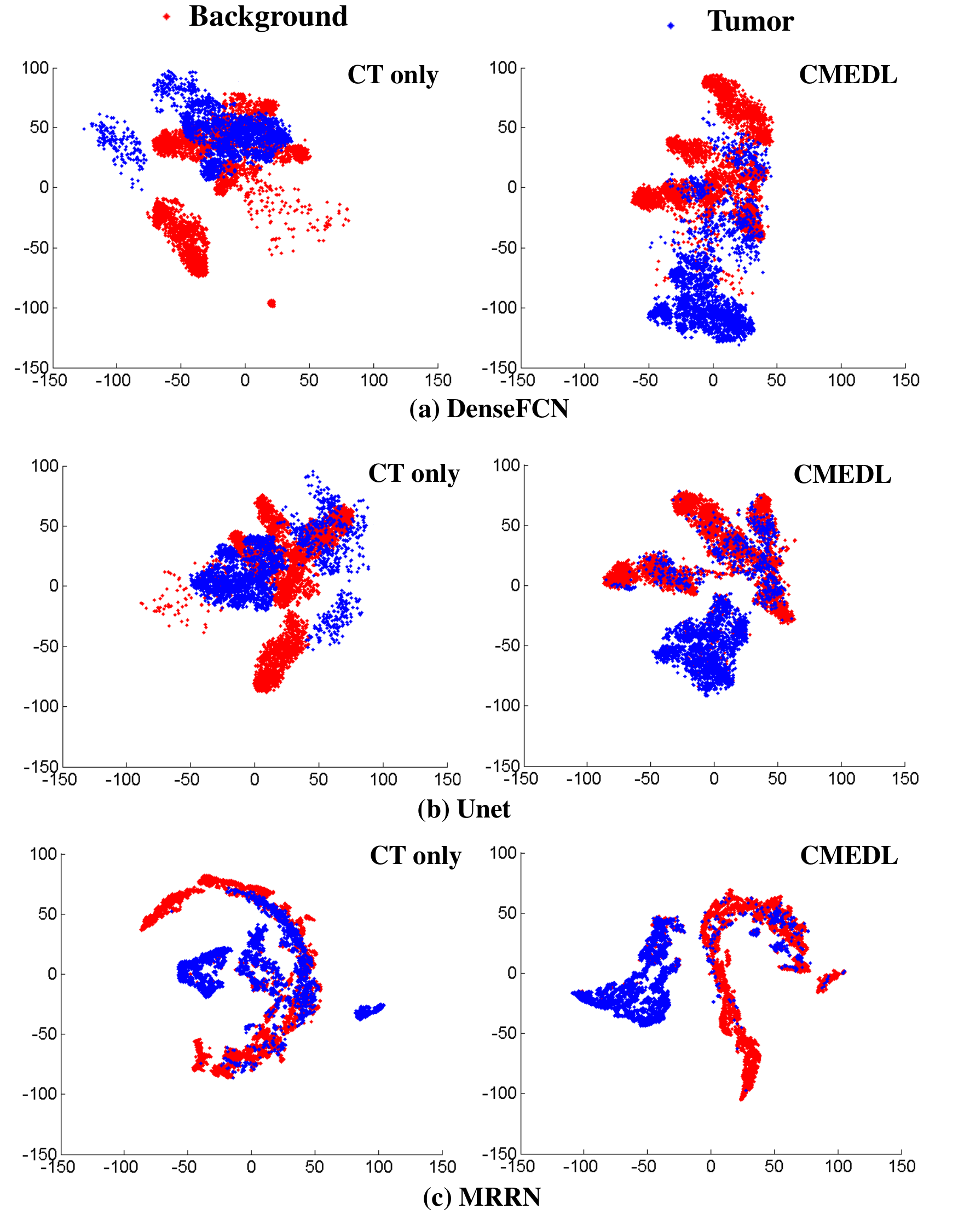}
		\vspace{-0.05cm}\setlength{\belowcaptionskip}{-0.4cm}\setlength{\abovecaptionskip}{0.08cm}\caption{\label{fig:tsne} \small \textcolor{black}{T-SNE map of the CMEDL CT vs. CT only features (last two layers) for (a) DenseFCN, (b) Unet, and (c) MRRN networks.} The T-SNE results clearly show that the CMEDL features better emphasize the difference between foreground and background.}
		\end{center}
\end{figure}

\begin{figure}[htb]
		\begin{center}
			\includegraphics[width=0.65\columnwidth,scale=1]{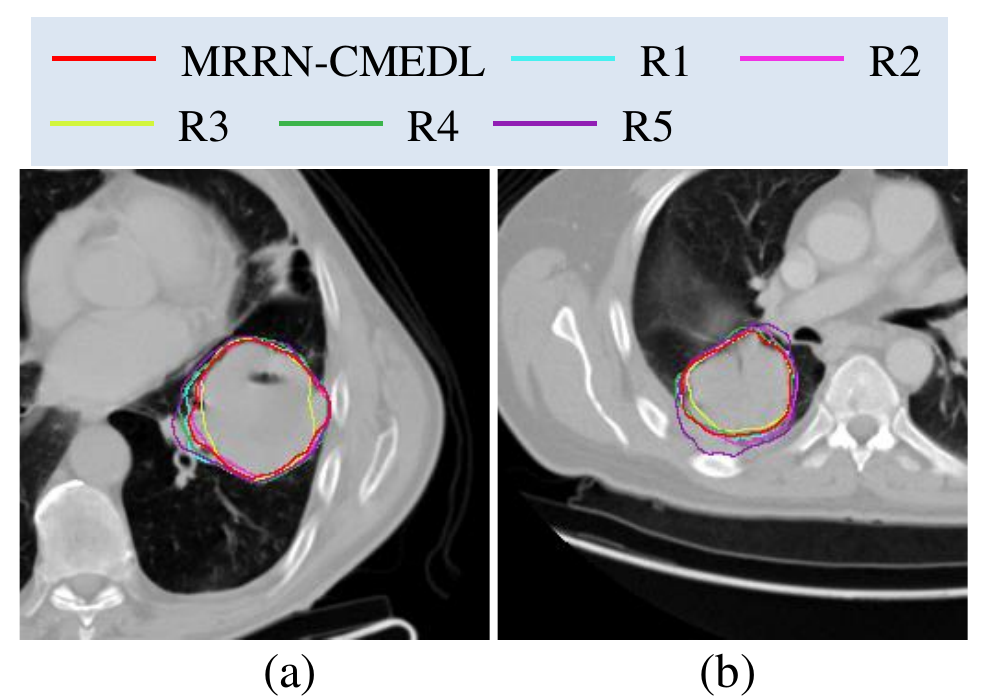}
			\vspace{-0.05cm}\setlength{\belowcaptionskip}{-0.4cm}\setlength{\abovecaptionskip}{0.08cm}\caption{\label{fig:iter_rater} \small \textcolor{black}{CMEDL- (red) and five radiation oncologists segmentation for two representative tumors. MRRN-CMEDL segmentation is closer to the average of the raters, while R3 and R4 tended to show under and over-segmentation, respectively.}}
		\end{center}
\end{figure} 				
\subsection{Effectiveness of CMEDL extracted features for separating tumor from background}
	We performed unsupervised clustering of the features extracted from the last two layers of \textcolor{black}{all CMEDL vs. non-CMEDL} networks using t-Stochastic Network Embedding (t-SNE) \cite{van2008visualizing} using the test dataset to study the effectiveness of the extracted features for differentiating tumor from background. Balanced number of tumor and background features (clipped to a total of 35,000 pixels per case) were extracted from within a 160$\times$160 patch enclosing the tumor in each slice containing the tumor, and input to t-SNE. The clustering parameters, namely perplexity was set at 60 and the number of gradient descent iterations was set to 1000.
\par
Features extracted using CMEDL networks provided better separation of tumor and background pixels then nonCMEDL networks(Fig. \ref{fig:tsne}). MRRN-CMEDL produced the best separation of tumor and background pixels (Fig.\ref{fig:tsne}(c)).

Fig.~\ref{fig:feature_map} shows visualization of feature maps in the channels one to twenty four of the last layer(with size of 256$\times$256$\times$64) for a standard Unet (Fig.~\ref{fig:feature_map}(a)), CMEDL Unet teacher network (Fig.~\ref{fig:feature_map}(b)), and the CMEDL CT student network (Fig.~\ref{fig:feature_map}(c)). As shown, the feature maps for the student network match the teacher network's activations very closely and better differentiate the tumor from its background compared to the CT only network. 
			\begin{figure*}[htp]
				\begin{center}
					\includegraphics[width=1.4\columnwidth,scale=1]{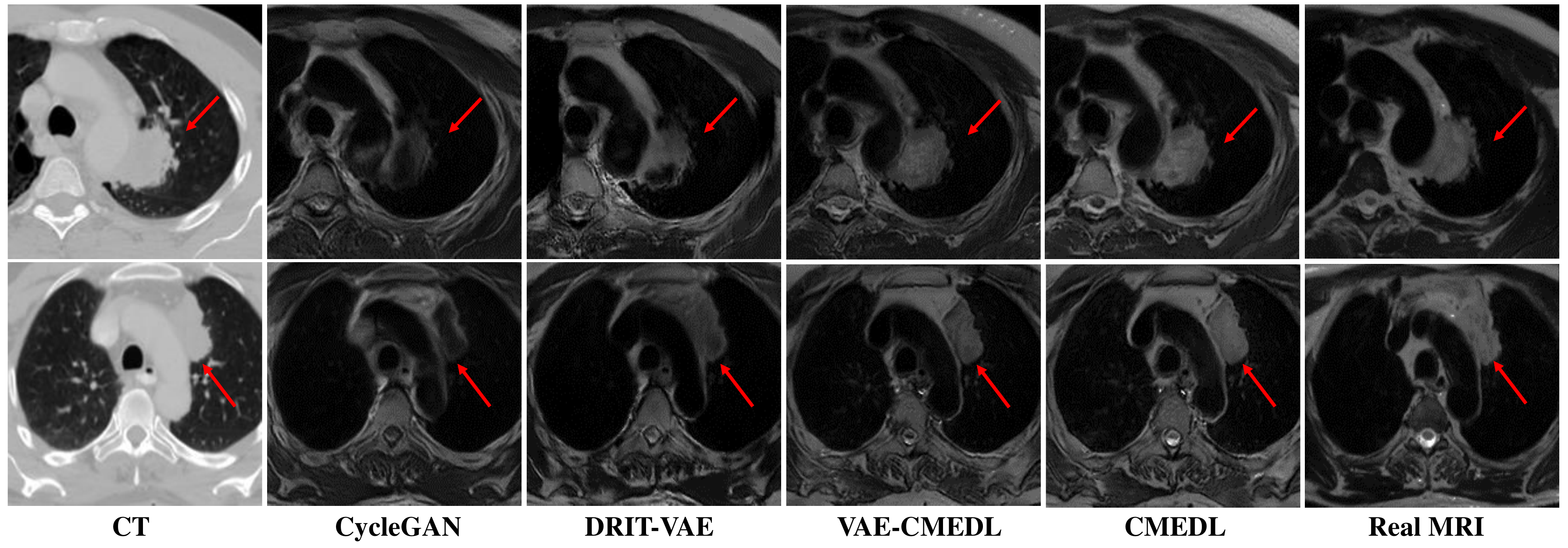}
					\vspace{-0.05cm}\setlength{\belowcaptionskip}{-0.4cm}\setlength{\abovecaptionskip}{0.08cm}\caption{\label{fig:Synthetic_result} \small \textcolor{black}{Representative examples of pMRI images  translated from CT images using CycleGAN, DRIT-VAE, VAE-CMEDL and CMEDL. Red arrow indicates lung tumor.}}
				\end{center}
			\end{figure*}

\subsection{Segmentation robustness to multiple raters}
We studied the robustness of the most accurate MRRN-CMEDL method against five radiation oncologist segmentations of NSCLC from an open-source dataset\cite{kalendralis2020fair} consisting of 20 patients imaged prior to radiation therapy. Robustness was measured using the coefficient of variation ($CV = \frac{\sigma}{\mu}$), where $\sigma$ is the standard deviation and $\mu$ is the population mean for the DSC and HD95 metrics. 
\par
MRRN-CMEDL had similar accuracy as all other raters, with a slightly higher DSC than when using R2 and R5 as reference (Table.~\ref{tab:Inter-observer}). It also showed lower coefficient of variation than all but R5 for HD95 and all but R1 and R4 for DSC accuracy. Fig.~\ref{fig:iter_rater} shows some representative cases with the CMEDL segmentation and the rater delineations. \textcolor{black}{Raters R4 and R3 showed larger variability in the segmentations than other raters. CMEDL on the other hand produced close to average segmentations.}	
			\begin{table} 
		\centering{
			\caption{Agreement with respect to multiple raters for MRRN-CMEDL. } 
			\label{tab:Inter-observer} 
			\centering
			\scriptsize
			\begin{tabular}{|c|c|c|c|c|c|c|} 
				\hline 
				
				\hline 
				{ Metric } & {  R1 } & {R2}  & {R3} & {R4} & {R5} & {CMEDL} \\
				
				\hline  
				
				\hline 
				{DSC } & 
				{0.846} & {0.805} & {0.824} & {0.832} & {0.810} & {0.825}\\
				{HD95(mm)} & {5.32} & {6.55} & {6.45} & {6.25} & {6.82} & {7.81}\\
				{CV$_{DSC}$} & {0.087} & {0.132} & {0.113} & {0.101} & {0.127} & {0.107}\\
				{CV$_{HD95}$} & {1.27} & {1.19} & {1.43} & {1.39} & {1.09} & {1.18}\\ 
				\hline   	
			\end{tabular}
		} 
\end{table}

\subsection{Pseudo MRI synthesis accuracy}
Synthesis accuracy was measured for CMEDL, VAE-CMEDL, cycleGAN only, and DRIT-VAE methods. PSNR and SSIM was also computed for 11 patients who had corresponding CT and MRI. 
\par
Standard CMEDL produced more accurate pMRI synthesis when compared with CycleGAN and the DRIT-VAE methods (Table.~\ref{tab:Img_translation_Acc}). CMEDL also produced more realistic synthesis of pMRI images compared to other methods as shown in Fig. \ref{fig:Synthetic_result}.
			
\begin{table}[h] 
		\centering{
			\caption{\textcolor{black}{Image translation accuracy on the lung dataset. Default CMEDL uses cycleGAN I2I network with contextual loss. } } 
			\label{tab:Img_translation_Acc} 
			\centering
			\scriptsize
			\begin{tabular}{|c|c|c|c|} 
				\hline 
				
				\hline 
				{ Method } & {  KL ($\downarrow$)} & {SSIM ($\uparrow$)}  & {PSNR ($\uparrow$)} \\
				
				\hline  
				
				\hline 
				{CycleGAN } & {0.34} & {0.60$\pm$0.03} & {14.05$\pm$1.04} \\
				{\textcolor{black}{DRIT-VAE}} & {0.22} & {0.74$\pm$0.02} & {17.38$\pm$1.56}\\
				{\textcolor{black}{VAE-CMEDL}}& {0.10} & {0.78$\pm$0.02} & {19.08$\pm$0.97} \\
				{CMEDL} & {0.079} & {0.85$\pm$0.02} & {20.67$\pm$0.93} \\

				\hline   	
			\end{tabular}
		} 
\end{table}

\subsection{Distillation learning with weak and equal teacher}
\begin{table}[b]
		\centering{
			\caption{\textcolor{black}{Weak teacher (CT) distillation for T2w-MRI segmentation. $^{*}$ indicates significant difference (p $<$ 0.001)}} 
			\label{tab:weak teacher} 
			\centering
			\scriptsize
			\footnotesize
			\begin{tabular}{|c|c|c|c|} 
				\hline 
				
				\hline 
				{ Method } & {  DSC ($\uparrow$)} & {HD95 mm ($\downarrow$)}  & {sDSC ($\uparrow$)} \\
				
				\hline  
				
				\hline 
				{MRI only } & {0.81$\pm$0.20$^{\star}$} & {6.24$\pm$6.50$^{\star}$} & {0.83$\pm$0.23$^{\star}$} \\
				{CMEDL} & {0.81$\pm$0.19}$^{\star}$ & {5.87$\pm$5.92$^{\star}$} & {0.84$\pm$0.22$^{\star}$}\\
				
				{CMEDL+pMRI} & {0.84$\pm$0.16} & {5.24$\pm$5.57} & {0.86$\pm$0.21} \\

				\hline   	
			\end{tabular}
		} 
	\end{table}

\begin{figure}
	\begin{center}
		\includegraphics[width=0.8\columnwidth,scale=1]{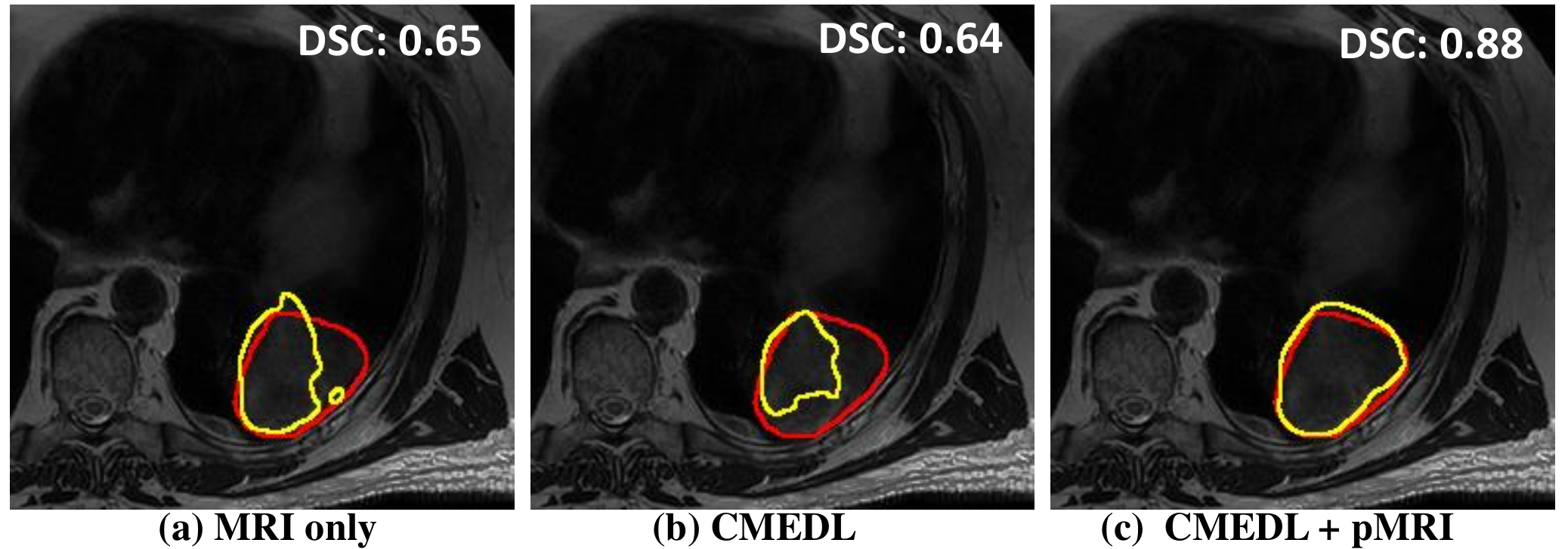}
		\vspace{-0.05cm}\setlength{\belowcaptionskip}{-0.4cm}\setlength{\abovecaptionskip}{0.08cm}\caption{\label{fig:mri_tumor_seg} \small \textcolor{black}{MRI tumor segmentation using (a) MRI only, CMEDL optimized with (b) hint loss only and (c) hint loss and pMRI augmented data. Red is the manual contour and yellow is the algorithm segmentation.  }}
    \end{center}
\end{figure}

\paragraph{Uninformative or weak teacher} \textcolor{black}{We studied the applicability of our approach \textcolor{black}{when performing distillation with a weak teacher} by using CT as the teacher modality to segment lung tumors on MRI (CT-MR\textcolor{black}{I} dataset). Distillation learning was done under two settings of: (i) hint losses only and (ii) hint losses combined with \textcolor{black}{student modality data augmentation with pMRI, similar} to other prior works\cite{Li_Yu_Wang_Heng_2020,dou2020}. MRRN-CMEDL network was trained with 3-fold cross validation. As shown in Table.~\ref{tab:weak teacher}, hint losses alone were insufficient to produce  accuracy improvement. On the other hand, combining hint losses with pMRI as augmented datasets (CMEDL+pMRI) showed significant accuracy improvement over both MRI only (\textcolor{black}{p \textless 0.001}) and CMEDL (\textcolor{black}{p \textless 0.001}) methods. \textcolor{black}{Fig~\ref{fig:mri_tumor_seg} shows a representative case with segmentations using the various methods. }}

\begin{table*} [htb]
\centering{\caption{\small{\textcolor{black}{MRRN-CMEDL accuracy using equal teacher distillation. Liver-LV, Spleen-SP, Left kidney-LK, Right kidney-RK. Overall average (Avg) is also shown. $^{*}$ indicates significant difference (p $<$ 0.05).}}} 
	\label{tab:adobmen_T1_T2_merge} 
	\setlength\tabcolsep{1 pt}
	\centering
	\scriptsize
	\centering
\begin{tabular}{c|c|c|c|c|c|c|c|c|c|c|c|c|c|c|c|c|c|c|c|c|c} 
			\hline
			
			\hline
	\multirow{3}{*}{Method}&\multirow{3}{*}{}&
	 \multicolumn{10}{c}{T2w MRI as Teacher, T1w MRI as Student
	 } & \multicolumn{10}{|c}{T1w MRI as Teacher, T2w MRI as Student}\\ 	 
	\cline{3-22}
	{}&{}  &  \multicolumn{5}{c}{DSC ($\uparrow$)} & \multicolumn{5}{|c}{HD95 mm ($\downarrow$)}& \multicolumn{5}{|c}{DSC ($\uparrow$)} & \multicolumn{5}{|c}{HD95 mm ($\downarrow$)}\\ 
	\cline{3-22}
	{} &{}& {  \textcolor{white}{A}LV }  & {  \textcolor{white}{A}LK  }& {  \textcolor{white}{A}RK }  & {  \textcolor{white}{A}SP  } & {\textcolor{white}{A}\textcolor{black}{Avg.} }
	& {  \textcolor{white}{A}LV }  & {  \textcolor{white}{A}LK  }& {  \textcolor{white}{A}RK }  & {  \textcolor{white}{A}SP  } & {\textcolor{white}{A}\textcolor{black}{Avg.}  }
	& {  \textcolor{white}{A}LV }  & {  \textcolor{white}{A}LK  }& {  \textcolor{white}{A}RK }  & {  \textcolor{white}{A}SP  } & {\textcolor{white}{A}\textcolor{black}{Avg.}  }
	& {  \textcolor{white}{A}LV }  & {  \textcolor{white}{A}LK  }& {  \textcolor{white}{A}RK }  & {  \textcolor{white}{A}SP  } & {\textcolor{white}{A}\textcolor{black}{Avg.}  }\\ 
\hline
	\hline
	\multirow{2}{*}{\textcolor{black}{MRRN}}&{Avg.} &{0.93\textcolor{black}{*}}&{0.86\textcolor{black}{*}}&{0.87\textcolor{black}{*}}&{0.86\textcolor{black}{*}}&\multirow{2}{*}{\textcolor{black}{0.88}}&{9.41}&{5.81}&{6.69}&{8.41}&\multirow{2}{*}{\textcolor{black}{7.58}}
	&{0.94}&{0.93}&{0.92}&{0.89\textcolor{black}{*}}&\multirow{2}{*}{\textcolor{black}{0.92}}&{7.9\textcolor{black}{*}}&{3.78}&{3.86}&{7.60}&\multirow{2}{*}{\textcolor{black}{5.79}}\\
	\cline{2-6}  \cline{8-11} \cline{13-16} \cline{18-21}  
{}&{Std.} &{	\textcolor{gray}{0.03}} &{	\textcolor{gray}{0.05}} &{	\textcolor{gray}{0.11}}&{	\textcolor{gray}{0.15}}&{}&{	\textcolor{gray}{8.27}}&{	\textcolor{gray}{3.84}}&{	\textcolor{gray}{7.30}}&{	\textcolor{gray}{5.08}}&{}

&{ \textcolor{gray}{0.02} }&{\textcolor{gray}{0.02}}&{\textcolor{gray}{0.04}}&{\textcolor{gray}{0.07}}&{}&{	\textcolor{gray}{5.80}}&{	\textcolor{gray}{2.70}}&{	\textcolor{gray}{2.09}}&{	\textcolor{gray}{5.94}}
\\
	\cline{1-22}
\multirow{2}{*}{\textcolor{black}{MRRN-CMEDL}}&{Avg.} &{0.94}&{0.89}&{0.90}&{0.88}&\multirow{2}{*}{\textcolor{black}{0.90}}&{7.59}&{5.09}&{5.01}&{7.05}&\multirow{2}{*}{\textcolor{black}{6.19}}

	&{0.94}&{0.94}&{0.93}&{0.90}&\multirow{2}{*}{\textcolor{black}{0.93}}&{6.56}&{3.30}&{3.43}&{6.81}&\multirow{2}{*}{\textcolor{black}{5.02}}

	\\
		\cline{2-6}  \cline{8-11} \cline{13-16} \cline{18-21}  
{}&{Std.} &{	\textcolor{gray}{0.02}}&{	\textcolor{gray}{0.07}}&{	\textcolor{gray}{0.08}}&{	\textcolor{gray}{0.04}}&{}&{	\textcolor{gray}{4.63}}&{	\textcolor{gray}{3.26}}&{	\textcolor{gray}{3.43}}&{	\textcolor{gray}{4.02}}&{}

&{\textcolor{gray}{0.02}   }& {\textcolor{gray}{0.02} }& {\textcolor{gray}{0.02}  }& {\textcolor{gray}{0.07}} &{} & {\textcolor{gray}{5.15}  } &{\textcolor{gray}{2.10} } &{\textcolor{gray}{1.47}  } & {\textcolor{gray}{3.26}}&{}\\	

	\hline
	
	\hline
	
	\end{tabular}} 
\end{table*}

	\begin{table} [b]
	        \centering{\caption{\small{Impact of each loss used in CMEDL}} 
		    \label{tab:ablation} 
		\setlength\tabcolsep{1 pt}
		\centering
		\scriptsize
		
		\centering
		\begin{tabular}{c|c|c|c|c|c} 
		    
			
			\hline
			
			\hline
			\multirow{2}{*}{Setting}&\multirow{2}{*}{$L_{cx}$}&\multirow{2}{*}{$L_{cyc}$}&\multirow{2}{*}{$L_{seg}^{rM}$}&\multirow{2}{*}{$L_{seg}^{pM}$}& \multirow{2}{*}{DSC}\\ 
			{}&{}&{}&{}&{}&{}\\ 
			
			\hline
			\hline	
			
			{1)}&{$\times$}&{$\checkmark$}&{$\checkmark$}&{$\checkmark$} &{0.72$\pm$0.19}\\
            \cline{1-6}

			\hline
			

			{2)}&{$\checkmark$}&{$\times$}&{$\checkmark$}&{$\checkmark$}&{0.71$\pm$0.19}\\

			\hline
			\hline

			{3)}&{$\checkmark$}&{$\checkmark$}&{$\times$}&{$\checkmark$}&{0.71$\pm$0.21}\\			
			\hline		
	
            {4)}&{$\checkmark$}&{$\checkmark$}&{$\checkmark$}&{$\times$}&{0.71$\pm$0.19}\\
            
            \hline
            {5)}&{$\checkmark$}&{$\checkmark$}&{$\checkmark$}&{$\checkmark$}&{0.75$\pm$0.17}\\
            
            \hline

			\hline
	
		\end{tabular}} 
	\end{table}	
\paragraph{Equal teacher}
\textcolor{black}{We also studied whether accuracy improvement can be reached through modality distillation between different image contrasts from the same modality (e.g. T1w MRI to T2w MRI and vice versa) for multi-organ segmentation. In this setting, both teacher and student modality contain similar amount of information in terms of visualization of the underlying anatomy.} 
\begin{figure}
	\begin{center}
	    \includegraphics[width=0.65\columnwidth,scale=1]{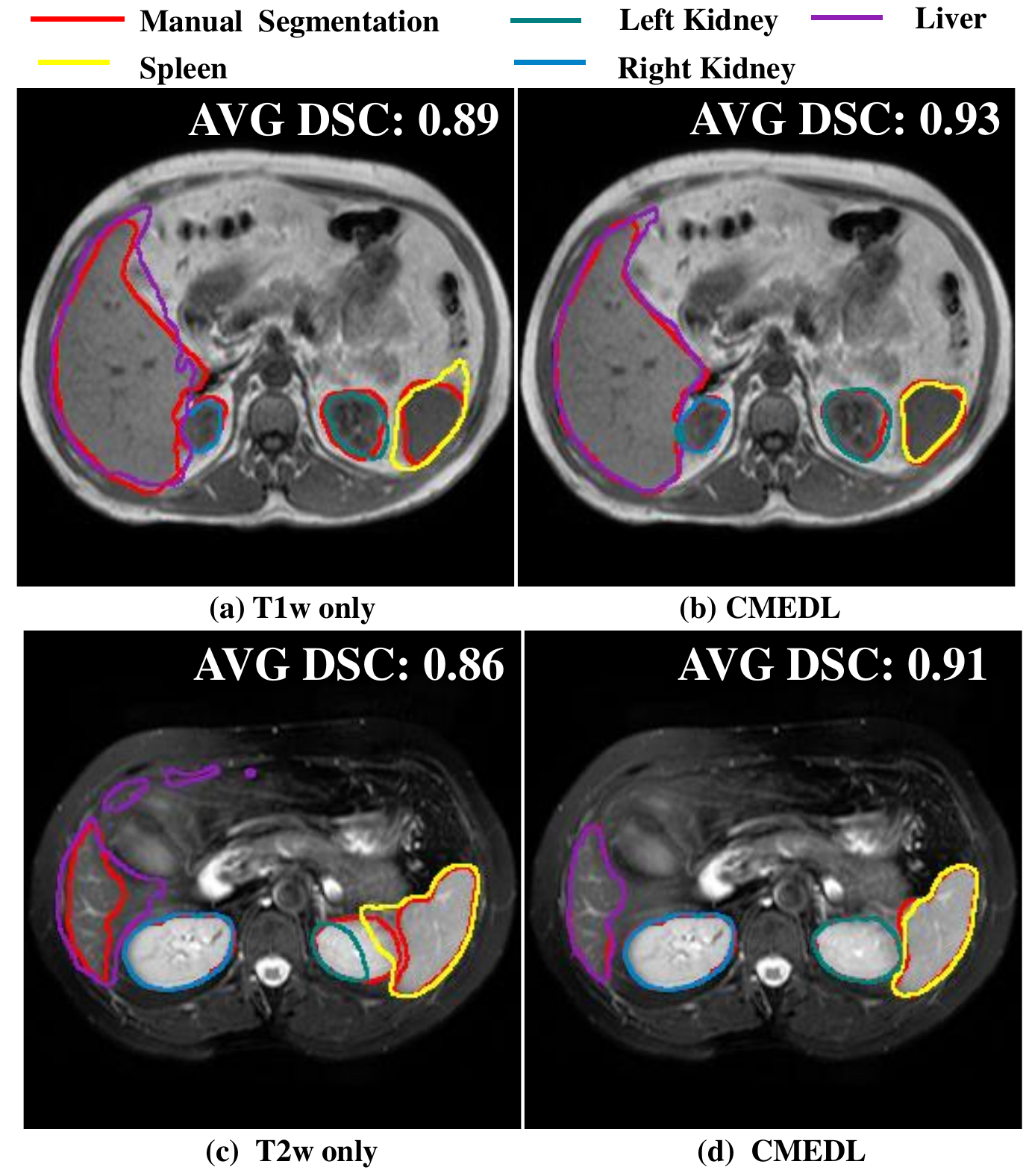}
	\vspace{-0.05cm}\setlength{\belowcaptionskip}{-0.4cm}\setlength{\abovecaptionskip}{0.08cm}
	\caption{\label{fig:t1w_t2w_overlay} \small \textcolor{black}{Segmentations for two representative cases on T1w and T2w by MRRN and MRRN-CMEDL. }}
	\end{center}
				
\end{figure}

\textcolor{black}{Fig.~\ref{fig:t1w_t2w_overlay} shows segmentations on two representative cases. Table.~\ref{tab:adobmen_T1_T2_merge} shows the segmentation accuracies for both T2w and T1w MRI trained with T1w and T2w MRI as teacher modalities, respectively. Whereas a \textcolor{black}{clear improvement was achieved from T1w MRI with T2w MRI as teacher for the left and right kidneys, CMEDL did not improve accuracy from T2w MRI, possibly due to the higher contrast on T2w MRI than T1w MRI for these organs}}.  
\subsection{Ablation experiments}
Ablation experiments were performed using the default CMEDL network (Unet segmentation and the cycleGAN I2I network). Segmentation accuracy is reported for the student network.
\paragraph{Impact of various losses} Impact of each loss, namely, contextual and cycle loss for the I2I network, as well as loss computed for the teacher network with augmented pMRI data was computed. Table.~\ref{tab:ablation} shows the accuracy with the removal of each loss. As shown, CMEDL accuracy decreased when either the labeled data from real MRI ($L_{seg}^{rM}$) or the augmented pMRI ($L_{seg}^{pM}$) was removed from the teacher network's training. However, the resulting accuracy was still better than a CT only network, indicating that accuracy gains are possible even with limited number of labeled examples for the teacher. Accuracy was also impacted by the removal of cycle consistency loss ($L_{cyc}$) and to a lesser extent from the contextual loss ($L_{cx}$) in the training of the I2I network, indicating the higher importance of cycle loss for accuracy. 
\paragraph{Impact of hint loss} We tested the more commonly used knowledge distillation loss\cite{hinton2015distilling} , which computes soft cross entropy between the teacher and student networks' outputs. We \textcolor{black}{separately} trained both Unet and MRRN networks with this loss. The scaling parameter, temperature \textit{T} was selected using grid search and set to 0.5. This method resulted in lower DSC accuracy of 0.72 $\pm$ 0.18 for the Unet and 0.75 $\pm$ 0.15 for the MRRN than the default CMEDL method (Table.~\ref{tab:segmentationAcc_val_test_tumor_Unet}).

\paragraph{Impact of feature layers used for distillation}
We evaluated the how the hints from different feature layers impacted accuracy by using hints from low-level (first two convolution layers), mid-level (the bottleneck layer before upsampling), and high-level (default last and penultimate) features. As shown, hints from the low-level features led to the worst accuracy (DSC of 0.69 $\pm$ 0.21). This accuracy is comparable to the non-CMEDL Unet method, indicating that forcing similar activations of the low-level features are not meaningful. Accuracy was slightly improved when using mid-level feature hints (DSC of 0.70 $\pm$ 0.20). On the other hand, there was a clear and significant (p $<$ 0.001) accuracy improvement when using the default high-level features (DSC of 0.75 $\pm$ 0.17), where only the anatomical contextual features are aligned between the two modalities compared to both low and mid-level feature hints.

\section{Discussion}
	We developed and validated a new unpaired modality distillation learning method called cross-modality educed distillation (CMEDL/"C-medal") applied to CT and MRI segmentation. We implemented unpaired distillation learning segmentation using three settings of an informative teacher (MRI as teacher and CT as student), an uninformative teacher (CT as teacher and MR as student), and an equally informative teacher (different MRI contrasts). CMEDL segmentations were significantly more accurate than other current methods. \textcolor{black}{CMEDL was most beneficial when using an informative teacher and produced accuracy gains for uninformative teacher distillation when using teacher modality to provide pseudo datasets as additional data for training.} We demonstrated the flexibility of our framework by implementing it with three different segmentation networks of varying complexity and two different I2I networks. 
	\textcolor{black}{ROC analysis showed that the CMEDL methods were more accurate than CT only methods. Furthermore, fusion of MRI information into CT improved over CT only segmentation regardless of the fusion strategy.}
	\\
	\textcolor{white}{AA}Our cross-modality distillation approach builds on prior results that showed accuracy gains when using MRI information to enhance inference on CT, albeit with paired CT-MR training sets\cite{leiPMB2020,FuMedPhys2020}. Our motivation to use unpaired datasets was to make the approach applicable to general clinical image sets, where paired CT-MR image sets are unavailable. 
	\\
	\textcolor{white}{AA}A key problem when using unpaired cross-modality datasets is how to effectively glean useful information given the lack of pixel-to-pixel coherence and the existence of any relationship only in the semantic space. \textcolor{black}{Prior works such as\cite{dou2020} and \cite{Li_Yu_Wang_Heng_2020} solved this issue by either learning a compact representation using distillation losses from the pre-softmax features combined with modality-specific feature normalization or by using the teacher modality to provide pseudo student modality data for data augmentation. These approaches also showed bi-directional accuracy improvements for both modalities. We used a different concept for distillation, wherein, a teacher modality regularizes feature extraction from a  student modality using hint losses applied to specific intermediate feature layers. This approach produced significant accuracy gains without data augmentation for the student modality when the teacher is more informative in terms of tissue contrast than the student. However, when the teacher is less informative, we found data augmentation from teacher modality to provide significant accuracy improvement as shown in\cite{Li_Yu_Wang_Heng_2020}. Accuracy improvement without data augmentation using informative teacher results from the fact that distillation loss is able to extract useful features for driving inference. In the absence of an informative teacher, the availability of additional pseudo modality datasets leads to accuracy improvements. However, we only found small accuracy gains in the equal or same modality distillation.}
	\\
	\textcolor{white}{AA}We found that concurrent training of the two networks provided accuracy gains even when the number of labeled teacher MRI modality datasets was lower than the student CT modality datasets. The accuracy improved over CT only method even in the extreme scenario where only augmented pseudo teacher modality datasets were used for training the teacher network. Thus, our approach obviates the need for large labeled data for pre-training the teacher network as is required in knowledge compression methods\cite{hinton2015distilling,chen2017learning,romero2014fitnets,gupta2016cross}.     
	\\
	\textcolor{white}{AA} \textcolor{black}{Unlike predominant distillation learning methods that used output distillation\cite{hinton2015distilling,Kats2019,Li_Yu_Wang_Heng_2020,KangMICCAI2020,wang2019}, wherein the student network is forced to mimic the outputs of the teacher by computing cross-entropy losses, we used hint learning\cite{romero2014fitnets} to minimize feature differences between intermediate layers}. We used the Frobenius norm to compute hint losses because it provides a stronger regularization than distribution matching using KL-divergence\cite{dou2020}. Our rationale for using hint losses was because MRI, which has a better soft tissue contrast than CT can help to guide extract “good” features for distinguishing foreground from background structures. Also, relevant is that differences in tissue visualizations in CT and MRI has been shown to lead to contouring differences\cite{karki2017}. Hence, we hypothesized that output distillation might provide less accurate results. \textcolor{black}{Our analysis showed that hint losses produced a clear accuracy improvement over the standard output distillation with temperature scaling}.
	\\
	\textcolor{white}{AA} Similar to\cite{dou2020}, which showed that performing distillation using the pre-softmax layers was beneficial, we found that hint loss distillation using higher-level features between the two modalities produced the best accuracy. This is sensible because CT and MRI only share a relationship in the semantic space such as the spatial organization of the anatomic structures but not the lower-level features. Particularly, pixel-to-pixel coherence may not exist when using unpaired data for training.
	\\
	\textcolor{white}{AA}We also found that significant accuracy improvement resulted even for a deep segmentation network like the MRRN, indicating that CMEDL is an effective technique for both shallow and deep networks.  \textcolor{black}{Although deeper networks are computationally intensive and may require large datasets for training, this problem could be alleviated by using cross-modality augmentation shown to be effective by others\cite{Li_Yu_Wang_Heng_2020}. Similarly, DRIT-VAE is a more accurate I2I network than cycleGAN, even when it requires more parameters to perform gradient update in the training than cycleGAN. However, both methods produced similar synthesis accuracy when used in the CMEDL framework. Similar synthesis accuracy of the two I2I networks resulted from the availability of additional regularization in the CMEDL network and also the use of contextual loss for the cycleGAN network.}  
	\\
	\textcolor{white}{AA}\textcolor{black}{A limitation of our framework is the use of 2D instead of 3D, due to the inherent restriction of the contextual loss computation from a pre-trained 2D VGG network, as well as the GPU memory limitation to extend the computation measuring feature losses in a 3D region. We note that 2D network methodologies have still shown promising accuracies\cite{Li_Yu_Wang_Heng_2020,KangMICCAI2020} for medical images. Similar to prior approaches\cite{dou2020,Li_Yu_Wang_Heng_2020,KangMICCAI2020}, we also did not study cross-modality distillation between anatomic and functional modalities (e.g. T1W MRI and diffusion MRI), because it is not clear if sufficiently accurate I2I synthesis is possible between modalities that capture very different tissue characteristics.} Importantly, CMEDL tumor segmentations showed good performance compared to radiation oncologist segmentations with lower variability, indicating its potential for clinical settings. 
    \section{Conclusions}
	We introduced a novel unpaired cross modality educed distillation learning approach (CMEDL) for segmenting CT and MRI images by leveraging unpaired MRI or CT image sets. Our approach uses the teacher modality to guide the extraction of features that signal the difference between structure and background on the student network. Our approach showed clear performance improvement over multiple segmentation networks.  CMEDL is a practical approach to using unpaired medical MRIs, and is a general approach to improving CT image analysis.
	

\bibliographystyle{IEEEtran}
\bibliography{refs}

\end{document}